\documentclass[useAMS,usenatbib,fleqn]{mnras}
\usepackage{amssymb,amsmath,amsopn,graphicx,epsfig,subfigure, psfrag, natbib}
\usepackage{amsopn,epsfig,subfigure, psfrag, natbib}
\newcommand{\HI}{{H{\ \!\sc i}~}}
\title[Foreground predictions for OWFA]{Simulated predictions for \HI at $z =
  3.35$ with the Ooty Wide Field Array - I. Instrument and the foregrounds}

\author[Marthi et al.]{Visweshwar Ram
  Marthi$^{1}$\thanks{E-mail: vrmarthi@ncra.tifr.res.in}, Suman
  Chatterjee$^{2,3}$, Jayaram N
  Chengalur$^{1}$, \newauthor Somnath Bharadwaj$^{2,3}$\\
$^{1}$National Centre for Radio Astrophysics, Tata Institute of
  Fundamental Research, Post Bag 3, Ganeshkhind, Pune - 411 007, India\\
$^{2}$Department of Physics and Meteorology, Indian
Institute of Technology Kharagpur, Kharagpur - 721 302, India\\
$^{3}$Centre for Theoretical Studies, Indian
Institute of Technology Kharagpur, Kharagpur - 721 302, India}

\date{Accepted 2017 July 14. Received 2017 June 25; in original form 2016
  November 22}

\pubyear{2017}

\begin{document}
\label{firstpage}
\pagerange{\pageref{firstpage}--\pageref{lastpage}}
\maketitle

\begin{abstract}
Foreground removal is the most important step in detecting the
large-scale redshifted \ion{H}{I} 21-cm signal. Modelling foreground spectra is 
challenging and is further complicated by the chromatic response of the telescope.
We present a multi-frequency angular power
spectrum(MAPS) estimator for use in a survey for redshifted \ion{H}{I} 21-cm
emission from $z\sim3.35$, and demonstrate its ability to accurately
characterize the foregrounds. This survey will be carried out with the
two wide-field interferometer modes of the upgraded Ooty Radio
Telescope, called the Ooty Wide Field Array(OWFA), at 326.5 MHz. 
We have tailored the two-visibility correlation for OWFA to estimate the MAPS and
test it with simulated foregrounds. In the process, we
describe a software model that encodes the geometry and the details of the
telescope, and simulates a realistic model for the bright radio sky. This article presents simulations
which include the full chromatic response of the telescope, in addition to the
frequency dependence intrinsic to the
foregrounds. We find that the visibility correlation MAPS estimator recovers the
input angular power spectrum accurately, and that the instrument response to the
foregrounds dominates the systematic errors in the recovered foreground power
spectra.
\end{abstract}


%%%%%%%%%%%%%%%%%%%%%%%%%%%%%%%%%%%%%%%%%%%%%%%%%%

%%%%%%%%%%%%%%%%% BODY OF PAPER %%%%%%%%%%%%%%%%%%

\section{introduction}
Measuring the post-reionisation neutral hydrogen power
spectrum provides a new probe of the large scale structure of the
universe \citep{Bharadwaj2001b, Wyithe2009, Pritchard2012}, and would allow
constraining many important cosmological
parameters \citep{Wyithe2008, Loeb2008, Bharadwaj2009, Visbal2009, Chang2010,
  Seo2010, Bull2015, Padmanabhan2015}.  The surge of interest in 21-cm intensity
mapping over the last decade stems from the promise of constraining dark energy.
The expansion of the universe is thought to accelerate at late times due to dark energy. The
Baryon Acoustic Oscillation (BAO) has been proposed as a standard ruler
\citep{Eisenstein2005a}. Its measurement at multiple redshifts can measure the
rate of expansion of the universe over cosmic time and hence place constraints
on dark energy \citep{Wang2006, Chang2008}. More recently, 21-cm intensity
mapping has gained attention due to its potential to
constrain the neutrino masses \citep[e.g.][]{ Pritchard2008,
  Pritchard2009, Oyama2013, Villaescusa-Navarro2015b}.
A number of post-reionisation experiments are currently being planned: interferometric
experiments, like the
Canadian Hydrogen Intensity Mapping Experiment (CHIME; \citealt{Bandura2014}) and those with other 
telescopes such as the Ooty Wide Field Array (OWFA;\citealt{Prasad2011, Subrahmanya2017a}), the Baryon
Acoustic Oscillation Broadband and Broad Beam 
Array (BAOBAB; \citealt{Pober2013b}),  and
the Tianlai Cylinder Radio Telescope (CRT; \citealt{Chen2011}), as well as
single dish intensity mapping experiments like BINGO \citep{Battye2016} and GBT-HIM \citep[e.g.][]{Chang2010}. The goals of
these experiments are similar - to constrain cosmological parameters through a
measurement of the power spectrum of the redshifted \ion{H}{I} 21-cm 
signal and to measure the BAO peak through a measurement
  of the power spectrum of the density fluctuations. This article discusses the
OWFA effort.

\citet{Ali2014} have presented calculations of both the \ion{H}{I} signal and the
foregrounds expected for OWFA, considering the diffuse Galactic synchrotron and the
extragalactic radio source contribution. \citet{Bharadwaj2015} present
Fisher-matrix forecasts of the \ion{H}{I} signal for OWFA, where
they predict that a 5$\sigma$ detection of the amplitude of \ion{H}{I} signal is
achievable in $\sim$ 150 hours of integration, assuming that foregrounds have
been completely removed.

The 39-MHz observing band at OWFA, centred at 326.5 MHz, provides
access to the \ion{H}{I} signal from a window of $\Delta z \sim 0.52$ around $z \sim
3.35$. A statistical detection of the post-reionisation signal is fraught with challenges very
similar to those in low frequency radio epoch of reionisation (EoR) experiments, the most formidable
of which is dealing foreground emission \citep[see e.g.][]{Furlanetto2006,Ali2008}. 
Foregrounds are important at least for two reasons: (1) the astrophysical
foregrounds are many orders of magnitude brighter than the cosmological
signal, and (2) foregrounds interact with the instrument to produce spectral
signatures that can contaminate the \ion{H}{I} signal \citep[see
  e.g.][]{Vedantham2012, Thyagarajan2015a}.

At least two classes of techniques to treat foregrounds exist.
In foreground removal methods, the foregrounds are mathematically
modelled as spectrally smooth functions and subtracted in the image
space \citep{Morales2006, Jelic2008, Bowman2009,  Liu2009b, Chapman2012}
or from the visibilities \citep{McQuinn2006, Gleser2008,Liu2009a,
    Petrovic2011}. \citet{Ghosh2011a, Ghosh2011b}
fit smooth functions to the multi-frequency angular power spectrum at each
multipole to subtract the foregrounds, exploiting the fact that the \ion{H}{I} signal
decorrelates more rapidly with frequency \citep{Bharadwaj2001a, Bharadwaj2005}
 than the foregrounds. 
Foreground isolation, on the other hand \citep{Datta2010, Morales2012,
  Parsons2012b, Vedantham2012, Thyagarajan2013, Liu2014}, uses
only those regions of the observed $\mathbfit{k}-$space that can be identified
as being less affected by the foreground emission.

Although it is known that the major foreground components all have smooth 
spectral behaviour, allowing them to be distinguished 
from the cosmological 21-cm signal, the telescope's
chromatic response introduces spectral features in the 
foreground contribution to the observed signal. These effects, the details of which are
telescope-specific, pose a serious challenge
for any foreground removal technique. Simulations that incorporate a detailed
frequency dependent telescope model are crucial to quantifying these effects as well
validating any foreground removal technique. In this paper we present a
software emulator that describes a chromatic model for OWFA. The
emulator is used in conjunction with a foreground sky model to predict the
foregrounds expected in observations with OWFA.  The three dimensional power
spectrum $P(\mathbfit{k})$ of the redshifted 21-cm brightness temperature
fluctuations \citep[e.g.][]{Bharadwaj2005} is well suited to quantifying the \ion{H}{I}
signal which, to a good approximation, may be assumed statistically
homogeneous in the three dimensional space spanned by the two angular
coordinates and the frequency axis. This assumption, however, is not valid for
the foregrounds, where the statistics are quite different in the angular
and frequency directions. The issue is further complicated
by the telescope's chromatic response. The Fourier modes $\mathbfit{k}$ are no
longer the ideal basis, and the power spectrum 
$P(\mathbfit{k})$ is no longer the obvious choice to quantify the statistical 
properties of the measured sky signal. In this paper we propose using the
correlations between the visibilities to quantify the statistical 
properties of the measured sky signal. The relation between the 
visibility correlations and the power spectrum $P(\mathbfit{k})$ has been studied
in several earlier works \citep{Bharadwaj2001a,Bharadwaj2005,Ali2014} and is
well understood. The multi-frequency angular power spectrum (MAPS; 
\citealt*{Datta2007}) provides a technique to jointly characterize
the angular and frequency dependence of the observed sky signal. This is very
closely related to the visibility correlations \citep{Ali2008,Ali2014}, and we
also use this to quantify our simulated foreground predictions for OWFA. 
In what follows we distinguish between three different kinds of
  visibilities, viz.: (a) the
model visibilities, which are the Fourier transform of the primary beam-weighted sky brightness distribution
for the array geometry under consideration, (b) the measured visibilities, which are the
actual measurements of the beam-weighted sky made by the array, corrupted by the complex
antenna gains and (c) the true visibilities, which are obtained from the measured
visibilities after the process of calibration, which is a noisy version of the
model visibilities.

To begin with, we introduce OWFA, the instrument, in
Section~\ref{sec:OWFA}, and describe the emulator in Section~\ref{sec:CaSPORT}.
In Section~\ref{sec:foregrounds} we introduce the foreground
components and describe the steps in the simulation of the foreground fields.
The model visibilities are obtained for the simulated foregrounds, given the instrument
description. Section~\ref{sec:MAPS} introduces the multi-frequency angular 
power spectrum (MAPS) estimator and its construction from the visibility
correlation for OWFA, which we use to estimate the angular power spectrum from the 
true visibilities. Section~\ref{sec:results} presents the results of the 
simulation and power spectrum estimation. We follow them up with a
discussion of the properties of the estimator in Section~\ref{sec:discussion}. 

\section{The Ooty Wide Field Array - OWFA}\label{sec:OWFA}
\begin{figure}
\begin{center}
  \includegraphics[scale=1.2]{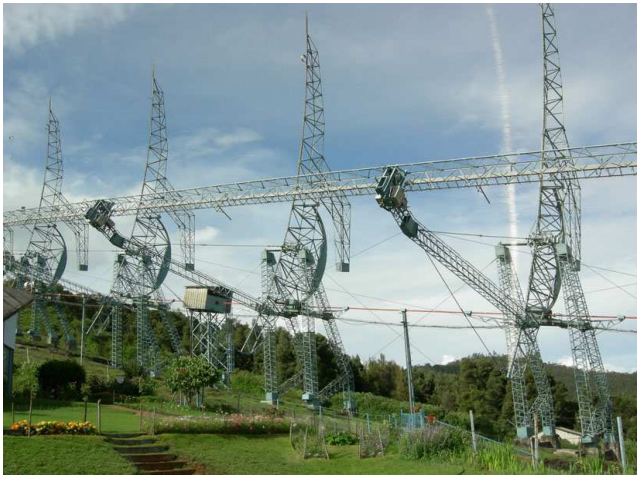}
  \caption{(colour online) A photograph of the Ooty Radio Telescope (ORT) on the
    sloping hill, showing four of the 22 parabolic frames. The left side of the
    picture is North and the right is South. }
  \label{fig:ORT-photo}
\end{center}
\end{figure}

\begin{figure}
\begin{center}
  \psfrag{x}{$x$}
  \psfrag{y}{$y$}
  \psfrag{z}{$z$}
  \psfrag{m}{$\mathbfit{m}$}
  \psfrag{n}{$\mathbfit{n}$}
  
\includegraphics[scale=0.295]{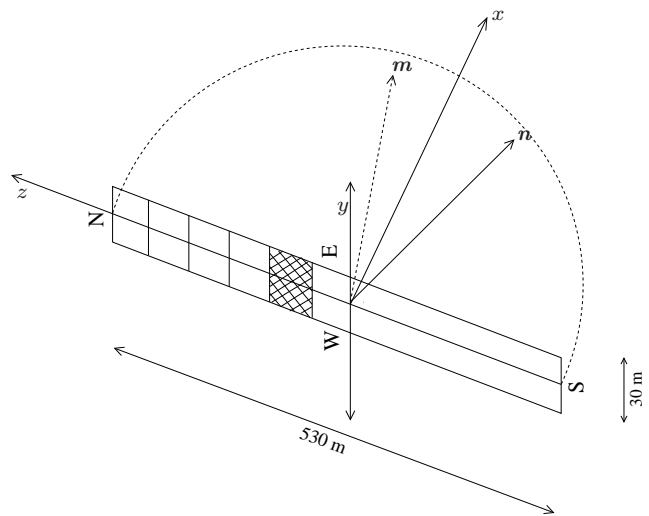}
\caption{This shows a  diagram of the $30 {\rm m} \, \times \, 530 {\rm m}$
aperture of the ORT  parabolic cylinder whose axis is alighned with the
N-S direction. The smaller rectangles, one of which is shaded, depcit
the apertures of the individual OWFA antennas which are linearly
arranged along the length of the ORT cylinder. The unit vector
$\mathbfit{n}$ denotes an arbitrary direction with coordinates
$(\alpha,\delta)$ on the celestial sphere. 
Our Cartesian coordinate system which is tied to the telescope has the
$z$-axis in the N-S direction, and the $x$-axis along the normal to the
aperture pointing towards the position $(\alpha_0,0)$ on the
celestial equator. The unit vector $\mathbfit{m}$ denotes the
phase center $(\alpha_0, \delta_0)$ which can be steered
electronically in the N-S direction.
}
\label{fig:owfa_coord}
\end{center}
\end{figure}

The Ooty Radio Telescope (ORT; \citealt{Swarup1971}) is located in the
Nilgiri Hills in the Indian peninsula, approximately at $11^\circ$N at
an altitude of $\sim 8,000$ ft.  
The telescope, operating at $326.5 \, {\rm MHz}$,  has an
offset-parabolic cylindrical reflector whose aperture is  a $530 \, {\rm  m}$ long and  $30 \, {\rm  m}$ wide
rectangle as shown in Figure~\ref{fig:owfa_coord}.
The feed consists of 1056 dipoles, each $0.5 \lambda$ long, all arranged
end to end along the length of the line focus of the reflector. 
 The telescope is located on a hill that is roughly sloped
$11^\circ$ North-South, which equals the latitude of the station. Figure
 ~\ref{fig:ORT-photo} shows a part of the ORT.
Effectively, the axis of the cylinder is parallel to the
earth's rotation axis, making it equatorially mounted. 
The telescope is steerable in the East-West direction mechanically by rotating 
the cylinder about its axis, allowing continuous tracking of a source on the sky
from rise to set. The telescope's response  is
steered electronically in the North-South direction by switching the phases of the
dipoles through an analog phase-switching network. Here 
we consider observations pointed towards the phase center
$(\alpha,\delta)=(\alpha_0, \delta_0)$, denoted by the unit vector $\mathbfit{m}$
(Figure~\ref{fig:owfa_coord}). 

%%%%%%%%%%%%%%%%

\begin{figure*}
\begin{center}
\includegraphics[scale=0.25]{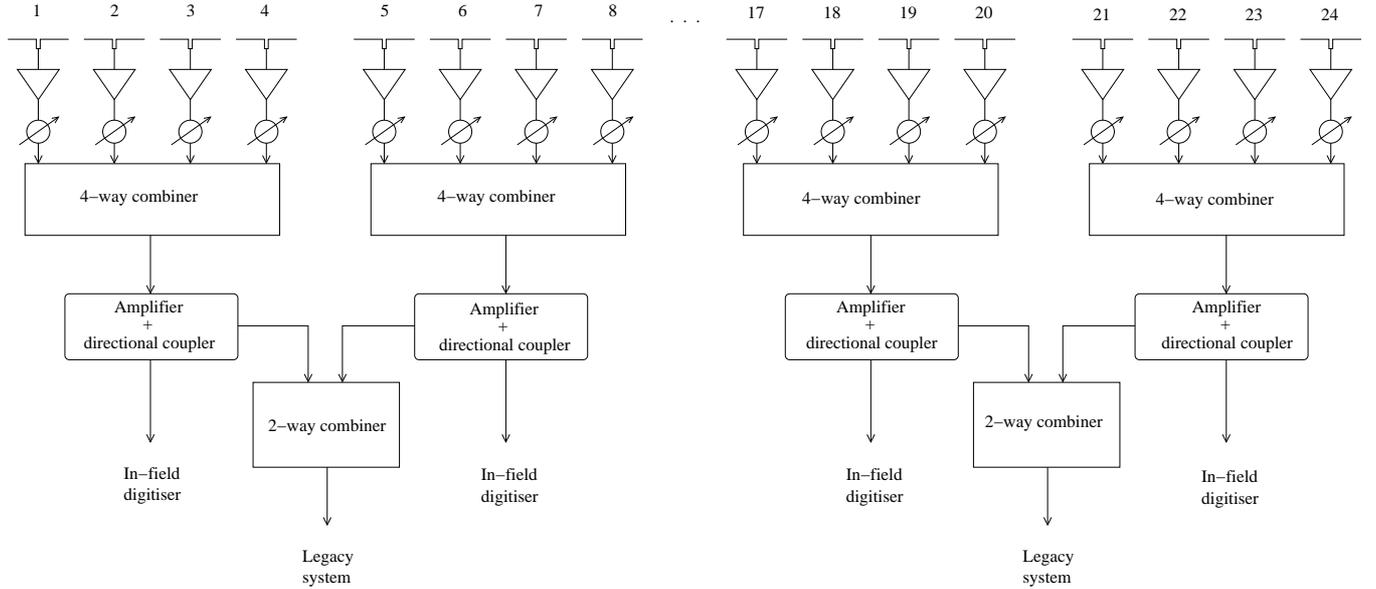}
\caption{The signal paths from the dipoles to the
  digitiser of OWFA: each dipole has an LNA and a phase shifter for
  setting the declination. The output of the 2-way combiner proceeds
  to the legacy ORT receiver system. The legacy ORT receiver system is a
  beam-former network capable of producing twelve simultaneous beams
  on the sky.}
\label{fig:sigchain}
\end{center}
\end{figure*}

In the legacy ORT system the signals from the dipoles are combined in a passive combiner 
network tree, only a part of which is shown Figure~\ref{fig:sigchain}. The
signals from $24$ successive dipoles are combined hierarchically in a series of
$4$-way, $2$-way and $3$-way combiners. The output at this stage is called a
half-module. Two half-module outputs are again combined in a $2$-way combiner:
therefore the $1056$ dipoles are grouped into $22$ modules of $48$ dipoles
each. The $22$ modules are split into $11$ Northern and $11$ Southern modules,
each half is used to generate $11$ phase-shifted beams, each pointing at a
different direction, and $1$ total intensity
beam. The beams obtained from the Northern and Southern modules are correlated with each other to get rid
of systematics. The legacy beam-former mode of the ORT system provides twelve
simultaneous beams on the sky.

The ORT is presently undergoing a major upgrade to operate as a programmable interferometer
\citep{Subrahmanya2017a}. The upgrade will enable two concurrent but
independent interferometer modes, besides retaining the workhorse
beam-former mode. The two concurrent modes are:
\begin{enumerate}
\item P-II : the signal is tapped after the $4$-way combiner - the
  second stage of the combiner network (Figure~\ref{fig:sigchain})
- to give a field-of-view (FoV) of
  $1.8^\circ\times27^\circ$ EW$\times$NS and a bandwidth of $\sim 39 \,
 {\rm MHz}$. It would operate as a linear array of $264$ equi-spaced antennas,
  giving $\sim 35,000$ baselines.
\item P-I: every group of six adjacent $4$-way combiner outputs are
  summed in phase in the software correlator system. This is
  equivalent to the signal being tapped at every half-module($24$
  dipoles), giving a FoV of $1.8^\circ \times
  4.5^\circ$ EW$\times$NS, and $\sim 39\, {\rm MHz}$ bandwidth. It would
  operate as an independent linear array of $40$ equi-spaced antennas,
  giving $780$ baselines.
\end{enumerate}

The RF signals tapped at the $4$-way combiner (Figure~\ref{fig:sigchain})
are digitised in the field
and the signals are transported to the central building on optical
fibre. The signals undergo two levels of pooling and a protocol
conversion before arriving in a high-performance cluster (HPC) that
will correlate the signals to obtain the visibilities. The P-I system
mentioned above will run in parallel with the P-II
system within the HPC.

OWFA will operate as an array of $N_A$ antennas arranged 
linearly along the length of the $530 \, {\rm m}$ ORT cylinder
as shown in Figure~\ref{fig:owfa_coord}. 
Each OWFA antenna has a $b \times d$ rectangular aperture, where $b=30 \, {\rm m}$ 
is the width of the parabolic reflector and $d$ is the length of the antenna which is 
different for P-I and P-II. $d$ is also the spacing between an adjacent antenna pair. 
The array parameters are summarized in Table~\ref{tab:OWFA}. 

The regular spacing of the antennas results in 
a large fraction of the $^{N_A}C_2$\footnote{$^{n}C_r$ is defined as $\frac{n!}{(n-r)!r!}$} baselines being redundant, and 
there are only $N_A-1$ unique baselines. We can express these unique
baselines in terms of the smallest baseline 
$\mathbfit{U}_1=d\ \! \nu / c$ as 
\begin{equation}
\mathbfit{U}_n = n \, \mathbfit{U}_1\, \, \, \, {\rm with } \, \, 1 \leq
n \leq N_A-1 \,.
\end{equation}  
Each baseline $\mathbfit{U}_n$ has a redundancy factor of $N_A - n$.

OWFA is conveniently described using a right-handed
Cartesian coordinate system tied to the telescope as shown in
Figure~\ref{fig:owfa_coord}. The $z$-axis is along the N-S
direction which is parallel to the axis of the parabolic cylinder, and the 
$x$-axis is normal to the telescope's aperture.
In this coordinate system the baseline vectors are all along the $z$-axis
and we have
\begin{equation}
\mathbfit{U}_n= |U_n| \mathbfit{z} \,.
\label{eqn:v-baselines}
\end{equation}
The unit vector $\mathbfit{n}$ towards an arbitrary direction
$(\alpha, \delta)$ on the celestial sphere can be expressed as 
\begin{equation}
\mathbfit{n}=\sin(\delta) \, \mathbfit{z} + \cos(\delta) [
  \cos(\alpha-\alpha_0) \, \mathbfit{x} + \sin(\alpha-\alpha_0)
  \mathbfit{y}] \,.
\end{equation}
where $\mathbfit{x}$, $\mathbfit{y}$ and
$\mathbfit{z}$ respectively denote the unit vectors along the $x$-, $y$-
and $z$-axes. The unit vector $\mathbfit{m}$ towards the phase center $(\alpha_0,\delta_0)$ 
can be represented as
\begin{equation}
\mathbfit{m}=\sin(\delta_o) \, \mathbfit{z} + \cos(\delta_0)
\, \mathbfit{x} \,.
\end{equation}

Observations at $326.5$ MHz relate to \ion{H}{I} at
  $z = 3.35$, which is at a comoving distance of $r= 6.84 \, {\rm
    Gpc}$. This relates each baseline $\mathbfit{U}$ to 
$k_\perp = \frac{2 \pi |\mathbfit{U}|}{r}$ which is the component of a 3D wave
  vector $\mathbfit{k}$  perpendicular to $\mathbfit{m}$.  
This sets the range 
$k^{\mathrm{min}}_\perp$ and $k^{\mathrm{max}}_\perp$ that will be probed by OWFA \citep{Ali2014}. 
Since OWFA is one-dimensional, it only probes $k_\perp$ along the length of
the interferometer.   
We also have $r' =|dr/d\nu|= 11.5 \, { \rm Mpc} \, {\rm MHz}^{-1}$ which sets 
the comoving interval $L=r' \, B_{\mathrm{bw}}$ spanned by $B_{\mathrm{bw}}$ the system
bandwidth of OWFA. This sets $k^{\mathrm{min}}_{\parallel}
  =\frac{2 \pi}{r' \,B_{\mathrm{bw}}}$ and $k^\mathrm{max}_{\parallel}  =\frac{\pi}{r' \, (\Delta\nu)}$ 
where $\Delta \nu$ is the channel width (frequency resolution) and $k_{\parallel}$ is the 
component of $\mathbfit{k}$  parallel to  $\mathbfit{m}$. 
Table~\ref{tab:OWFA} summarizes the important parameters of the two
interferometer modes, assuming the \emph{WMAP} 9-year results for the
cosmological parameters \citep{Hinshaw2013}.
\begin{table}
\centering
\caption{Parameters of OWFA. OWFA is a one-dimensional interferometer; the
  values for $k$ pertain only to the north-south extent of the array elements.}
\label{tab:OWFA}
\begin{tabular}{l||c|c}
\hline Parameter & P-I & P-II \\ \hline Antennas & 40 & 264 \\ Total
baselines & 780 & 34716 \\ Unique baselines & 39 & 263 \\ Shortest
baseline & 11.5 m & 1.92 m \\ Longest baseline & 448.5 m & 505.0
m\\ Central frequency & 326.5 MHz & 326.5 MHz \\ Bandwidth & 39 MHz &
39 MHz \\ Spectral resolution & 125 kHz & 125 kHz \\ Aperture ($b
\times d$) & $30 \times 11.5$ m$^2$ & $30 \times 1.97$ m$^2$ \\ FoV at
$\delta = 0^\circ$ & $1.8^\circ \times 4.5^\circ$ & $1.8^\circ \times
27^\circ$ \\ Resolution & $0.1^\circ\sec\delta_0 \times 1.8^\circ$ &
$0.1^\circ\sec\delta_0 \times 1.8^\circ$ \\ $k^{min}_\perp$ &
$1.1\times 10^{-2}\ \mathrm{Mpc}^{-1}$ & $2.0\times
10^{-3}\ \mathrm{Mpc}^{-1}$ \\ $k^{max}_\perp$ & $4.6\times
10^{-1}\ \mathrm{Mpc}^{-1}$ & $5.2 \times 10^{-1}\ \mathrm{Mpc}^{-1}$
\\ $k^{min}_\parallel$ & $1.4\times 10^{-2}\ \mathrm{Mpc}^{-1}$ &
$1.4\times 10^{-2}\ \mathrm{Mpc}^{-1}$ \\ $k^{max}_\parallel$ &
$4.6\ \mathrm{Mpc}^{-1}$ & $4.6\ \mathrm{Mpc}^{-1}$\\
%$k^{max}_\parallel$ &  $11.2\ \mathrm{Mpc}^{-1}$ &  $11.2\ \mathrm{Mpc}^{-1}$\\
\hline
\end{tabular}
\end{table}

\section{A software emulator for OWFA}\label{sec:CaSPORT}
In this Section, we will describe a software emulator for OWFA that we
subsequently use to make foreground predictions.  The
emulator (\textbf{Prowess}; \citealt{Marthi2017a}) was written (1) to simulate
the observed radio sky at 327 MHz, (2) to provide a 
pipeline to simulate visibilities, introduce extrinsic systematics,
study intrinsic systematics and allow us to
test calibration schemes \citep[see][]{Marthi2014} and detection strategies even
as the telescope undergoes an upgrade, (3) to prepare a
science-ready observatory data analysis software pipeline in time for
first light. For the purpose of this article, only the first and second items are
relevant. We describe \textbf{Prowess}\footnote{\textbf{Prowess} is available at https://github.com/vrmarthi/prowess} 
and some of the details of its design briefly in this section. It is important to
note that the telescope model incorporated in the emulator includes 
the chromatic effects associated with the primary
beam and the baselines. 
\begin{figure*}
\centering
\begin{minipage}{190mm}
\subfigure[Aperture
  arrangement]{\label{subfig:antenna-geometry}\includegraphics[scale=0.2]{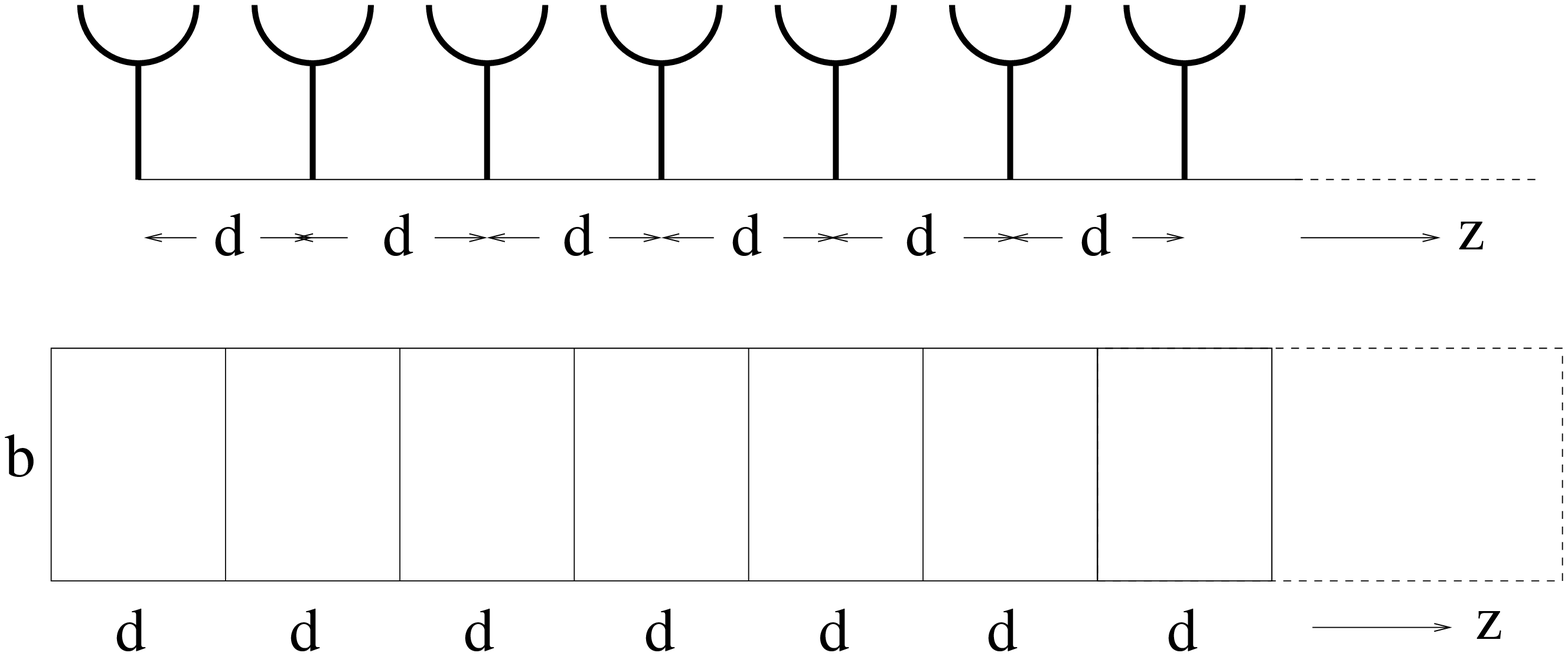}}
\hskip 4mm
\subfigure[Primary beam]{\label{subfig:pbeam}
\includegraphics[scale=0.2]{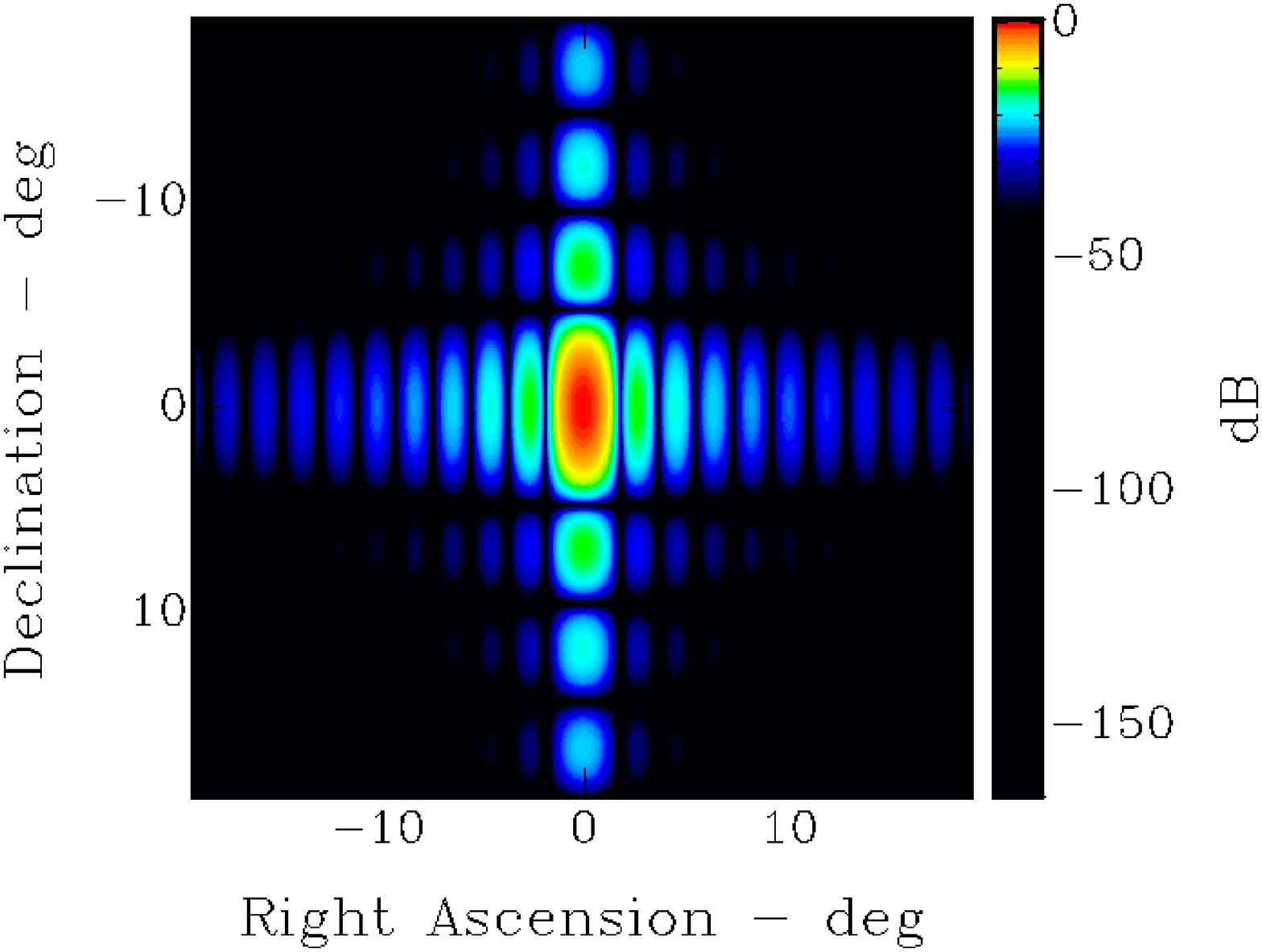}}
\hskip 4mm
\subfigure[Baseline
  space]{\label{subfig:baseline-space}\includegraphics[scale=0.22]{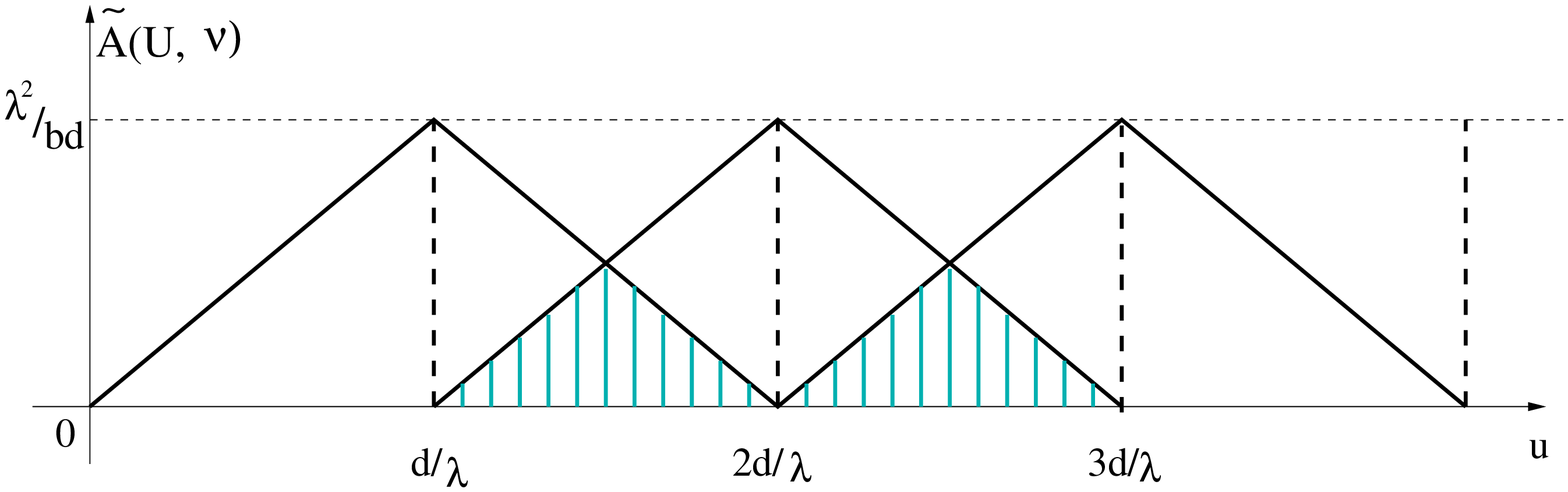}}
\end{minipage}
\caption{(colour online) The aperture arrangement for OWFA in real and baseline space.}
\end{figure*}

\subsection{Visibilities}
The visibilities $\mathbfit{M}(\mathbfit{U})$ are the Fourier
transform of the product of the specific intensity  
distribution on the sky $I(\mathbfit{n},\nu)$ and the telescope's primary beam 
pattern $A(\Delta \mathbfit{n},\nu)$ where $\Delta \mathbfit{n} =
\mathbfit{n} - \mathbfit{m}$ (Figure~\ref{fig:owfa_coord}). 
 The visibilities can be calculated as 
 \begin{equation} 
\mathbfit{M}(\mathbfit{U})=\int d \Omega_{\mathbfit{n}} \,
I(\mathbfit{n},\nu)  \,  A(\Delta \mathbfit{n},\nu) \,  
 e^{-2 \pi i \mathbfit{U} \cdot \Delta \mathbfit{n}} \,
\label{eqn:model_RIME}
\end{equation}
where the solid angle integral is over the entire celestial
hemisphere. 

Here $A(\Delta \mathbfit{n},\nu)$ is the  primary beam power pattern of OWFA,
modelled as a two-dimensional sinc-squared function
\begin{equation}
 A(\Delta \mathbfit{n},\nu)= {\rm sinc}^2 \left( \frac{\pi b \nu \, \Delta n_y}{c}
 \right) {\rm sinc}^2 \left( \frac{\pi d \nu \, \Delta n_z}{c}
 \right) 
\label{eq:pb}
\end{equation}
where $\Delta n_y$ and $\Delta n_z$ are respectively the $y$ and $z$
components of $\Delta \mathbfit{n}$. The primary beam pattern here is taken to be
the diffraction pattern of rectangular aperture of dimensions $b \times
d$, the values of $b$ and $d$ for OWFA are given in Table~\ref{tab:OWFA}. This
form of the primary beam has been assumed as it likely to lead to larger
systematics because of its high sidelobe levels and is hence
conservative. In reality, the tapered east-west illumination of the aperture
results in an approximately truncated Gaussian beam.
\textbf{Prowess} is capable of handling any user-defined primary beam.
Figure~\ref{subfig:antenna-geometry} shows the
aperture arrangement and Figure~\ref{subfig:pbeam} shows the P-I primary beam pattern as a function of
the celestial coordinates $\alpha$ and $\delta$ for a pointing towards $(\alpha_0,\delta_0) = (0,0)$.

Note that in this co-ordinate system the phase factor which is the 
argument of the exponent in equation (\ref{eqn:model_RIME})
\begin{equation}
2 \pi\, i\, \mathbfit{U} \cdot \Delta \mathbfit{n}=2 \pi\,i\, |\mathbfit{U}| [\sin(\delta)-\sin(\delta_0)]
\end{equation}
 depends only on the baseline length and the
declinations, reflecting the 1D geometry of OWFA.

\subsection{Overview of the software pipeline}
We begin with a geometric description of the array in an input Antenna
Definition file. Since the telescope is equatorially mounted, the baseline
co-ordinates remain stationary as the antennas
track the pointing coordinates on the sky. The baseline vectors are obtained from the physical antenna
separations defined at the central frequency but scaled appropriately at each channel
when computing the visibilities.
\begin{equation}
\mathbfit{d}_{|a-b|} = \mathbfit{x}_a - \mathbfit{x}_b,\ \ \ \  
\mathbfit{U}_{|a-b|} = \mathbfit{d}_{|a-b|}\ \frac{\nu}{c}
\label{eq:basln}
\end{equation}
It is worth noting that the chromatic response of the interferometer
enters the picture through equations (\ref{eq:pb}) and (\ref{eq:basln}) 
which respectively capture the fact that the primary beam pattern and the
baseline distribution both vary with frequency. Even for a perfectly achromatic source, the chromatic response of the
interferometer will introduce a frequency dependent structure in the 
measured visibilities. 

The pixel sizes in all our simulated maps and the primary
beam are $\sim 1 \arcmin \times 1\arcmin$. The maps and the primary beam
are all $2048 \times 2048$ pixels wide, spanning an angular extent of $\sim38^\circ$
in each dimension.
In this paper, we will focus only on P-I, where the antenna
primary beam  full width at half maximum (FWHM) is $1.8^\circ \times 4.5^\circ$. 

The brightness distribution for the diffuse foreground component in our
simulations is
obtained as a random realisation of its power spectrum (see 
Section~\ref{sec:foregrounds}). The specific intensities $I(\mathbfit{n}_p,\nu)$  of
the pixelized sky map are converted to flux  
densities by scaling with the appropriate solid angle 
\begin{equation}
  S(\mathbfit{n}_p, \nu)=I(\mathbfit{n}_p, \nu) \, \Delta \Omega \,,
  \label{eqn:I-to-S}
\end{equation}
the index $p$ here refers to the $N_P$ pixels in the map.  
The extragalactic point source foreground is
obtained as a random realisation of its power spectrum 
and obeying the differential source count relation (see Section~\ref{sec:foregrounds}).  
This results in a set of $N_S$ discrete extragalactic radio sources 
identified by their respective positions $\mathbfit{n}_s$
and flux densities $S(\mathbfit{n}_s,\nu)$.

We note that similar simulations of both the instrument and foregrounds
  have been carried out for CHIME \citep{Shaw2014} and MWA \citep{Thyagarajan2015a}.
In contrast to the above full sky foreground simulations we emphasize that for P-I of OWFA, where the
  primary beam is much smaller in comparison, a full
  sky treatment of the foregrounds is not necessary. However, unlike the small FoV LOFAR
foreground simulations \citep{Jelic2008}, we do consider a FoV
  much further out than the primary. For P-II, with a primary
  FoV of 0.5 rad, wide-field effects become important, and
  they will be considered separately for the P-II forecasts.

The model visibilities $\mathbfit{M}(\mathbfit{U})$ for the
non-redundant set of baselines are obtained using a 
discretised version of equation~(\ref{eqn:model_RIME}):
\begin{equation}
\mathbfit{M}(\mathbfit{U}) =
\sum\limits_{b} \ S(\mathbfit{n}_b,\nu)\ A(\Delta \mathbfit{n}_b,\nu) 
\ e^{-i 2\pi   \mathbfit{U} \cdot \Delta \mathbfit{n}_b} \,.
\label{eqn:modelvis}
\end{equation} 
The sum denoted here by the index  $b$ includes the $N_P$ pixels of the diffuse
foreground as well as the $N_S$ discrete extragalactic sources.

The visibility $\mathbfit{V}$ measured by a
baseline with antennas $a$ and $b$ depends on the model
visibility for that particular spacing, as well the element gains:
\begin{equation}
\mathbfit{V}\left(\mathbfit{U}_{ab} \right) =
g_a\ g_b^*\ \mathbfit{M}\left(\mathbfit{U}_{|a-b|} \right) +
\mathcal{N}\left(\mathbfit{U}_{ab} \right) 
\label{eqn:RIME}
\end{equation}
where $g_a$ and $g_b$ are the complex antenna
gains (which could also be frequency dependent)  and $\mathcal{N}$ is Gaussian
random noise equivalent to the system 
temperature $T_{\mathrm{sys}}$.  

The antenna gains and the true (calibrated) visibilities are estimated from the measured visibilities
by employing the non-linear least squares minimisation algorithm described in
\citet{Marthi2014}. Due to the enormous redundancy of the measurements, the
highly overdetermined system of equations enables simultaneously solving for
both the gains and the true visibilities. This is interesting particularly in
light of studies regarding the limitations introduced by calibration based on an
incomplete model of the sky, as is routinely employed in interferometry \cite{Patil2016}.
The real and
imaginary part of the noise $\mathcal{N}$ in equation~(\ref{eqn:RIME})
each has a RMS fluctuation \citep{thompson2008}
\begin{equation}
\sigma_{ab} = \frac{\sqrt{2}k_B T_{\mathrm{sys}}}{\eta A
  \sqrt{\Delta\nu\Delta t }}
\label{eqn:T_sys}
\end{equation}
per channel, where $k_{\mathrm{B}}$ is the Boltzmann constant, $\eta$ is the
aperture efficiency, $A = b \times d$ is the aperture area,
$\Delta\nu$ the channel width and $\Delta t$ the integration time. 
The foreground maps give the flux in Jy units at every
pixel, therefore the Fourier sum directly produces the model
visibilities in Jy units. The flowchart in Figure~7 of \citet{Marthi2017a} gives an
overview of the part of the emulator used to obtain the
visibilities. As mentioned earlier, the emulator fully incorporates
the chromatic response of the instrument. The baselines
(eq.~\ref{eq:basln}), the primary beam (eq.~\ref{eq:pb}) and 
the sky vary with frequency. These quantities were
calculated separately at each frequency and used in equation ~(\ref{eqn:modelvis}) to
calculate the visibilities. The spectral variation of the sky 
is discussed in Section \ref{sec:foregrounds}. The pixel solid angle
$\Delta\Omega$ in equation (~\ref{eqn:I-to-S}) is
held fixed across the band. In our simulations of the foregrounds, the sky is sampled
at a much higher resolution than the resolution element of OWFA.
 %The brightness distribution map for each of emission components and
 %the primary beam generated can optionally be written to disk and displayed.
 The simulated visibilities, the primary beam and the foreground maps are written to
disk in the Flexible Image Transport System (FITS; \citealt{FITS})
format. The visibility records are written to UVFITS files at one-second
interval. 

In \textbf{Prowess}, the observing band is centred at 326.5 MHz with a bandwidth of $\sim 39$ MHz split
into 312 channels in the simulations. The frequency resolution is thus
125 kHz per channel. Based on our understanding
of the distribution of the neutral gas around the redshift of $z\sim3.35$, the \ion{H}{I} signal at two redshifted frequencies, separated by more
than $\Delta\nu \sim 1$ MHz, is expected to decorrelate rapidly \citep{Bharadwaj2005,
  Bharadwaj2009, Chatterjee2017}. This means that with a channel
resolution of 125 kHz, the \ion{H}{I} signal correlation is adequately sampled over the
1-MHz correlation interval. In reality, the channel resolution is
likely to be much finer ($\sim 50 \mathrm{kHz}$), with about 800 channels across the
39-MHz band. This is useful for the identification and excision of narrow line
radio frequency interference. Beyond the need to handle RFI, there is no 
incentive to retaining the visibilities at this resolution. Eventually, we may smooth the
data to a resolution of 125 kHz, keeping in mind the decorrelation bandwidth of
the \ion{H}{I} signal. \textbf{Prowess} itself is indeed capable of running at any
frequency resolution, including the actual final configuration of the P-I and P-II
systems. But the 125-kHz resolution used in the simulations allows for rapid
processing, especially if they are to be run repetitively for a wide range of different parameters.
%
%%---------------------------------------------------------
\section{The foregrounds}\label{sec:foregrounds}
The most dominant contributors to the foregrounds are the diffuse synchrotron emission from our 
Galaxy and the extragalactic radio sources \citep[see e.g.][]{DiMatteo2002, Santos2005, Furlanetto2006, Ali2008}. Although some choice is available in 
the selection of a reasonably cold region of the Galaxy through which to observe the 
distant universe, emission from extragalactic radio sources is
ubiquitous.
At different angular scales, either of the two dominates: at the very large scales, 
$\ell \lesssim 100$, the Galactic synchrotron emission contributes more to the
foreground budget. At smaller scales, there is a 
gradual transition to the flux from extragalactic
sources. Galactic synchrotron is likely to dominate the emission
at the largest angular scales observable by P-II at the OWFA, and likely
at the first few baselines of P-I \citep{Ali2014} at higher
declinations. Since the largest contribution to the \ion{H}{I} signal is
expected to come from the largest angular scales \citep{Chatterjee2017} we would have access
to, the large scale emission from the diffuse foreground assumes importance.

\subsection{The Diffuse Galactic Foreground}
Diffuse synchrotron emission from the Galaxy has been studied
extensively, both within the context of statistical \ion{H}{I}
experiments \citep{Jelic2008, Bernardi2009, Jelic2014} as well as in their own
right \citep[e.g.][]{Haslam1981, Haslam1982}. \citet{LaPorta2008} determine the power spectrum of
the diffuse synchrotron emission at scales larger than 0.5$^o$ from
the 408-MHz Haslam total intensity maps, where they find that the
index of the power law $\gamma$ varies between 2.6 and 3.0. Similarly,
there are measurements at scales smaller than 0.5$^o$ from the
WSRT \citep{Bernardi2009, Bernardi2010}, GMRT\citep{Ghosh2012} and
LOFAR \citep{Iacobelli2013b}. 
\begin{figure*}
\centering
\begin{minipage}{190mm}
\subfigure[Simulated map]
{\label{fig:GAL-map}
\includegraphics[scale=0.3,angle=270]{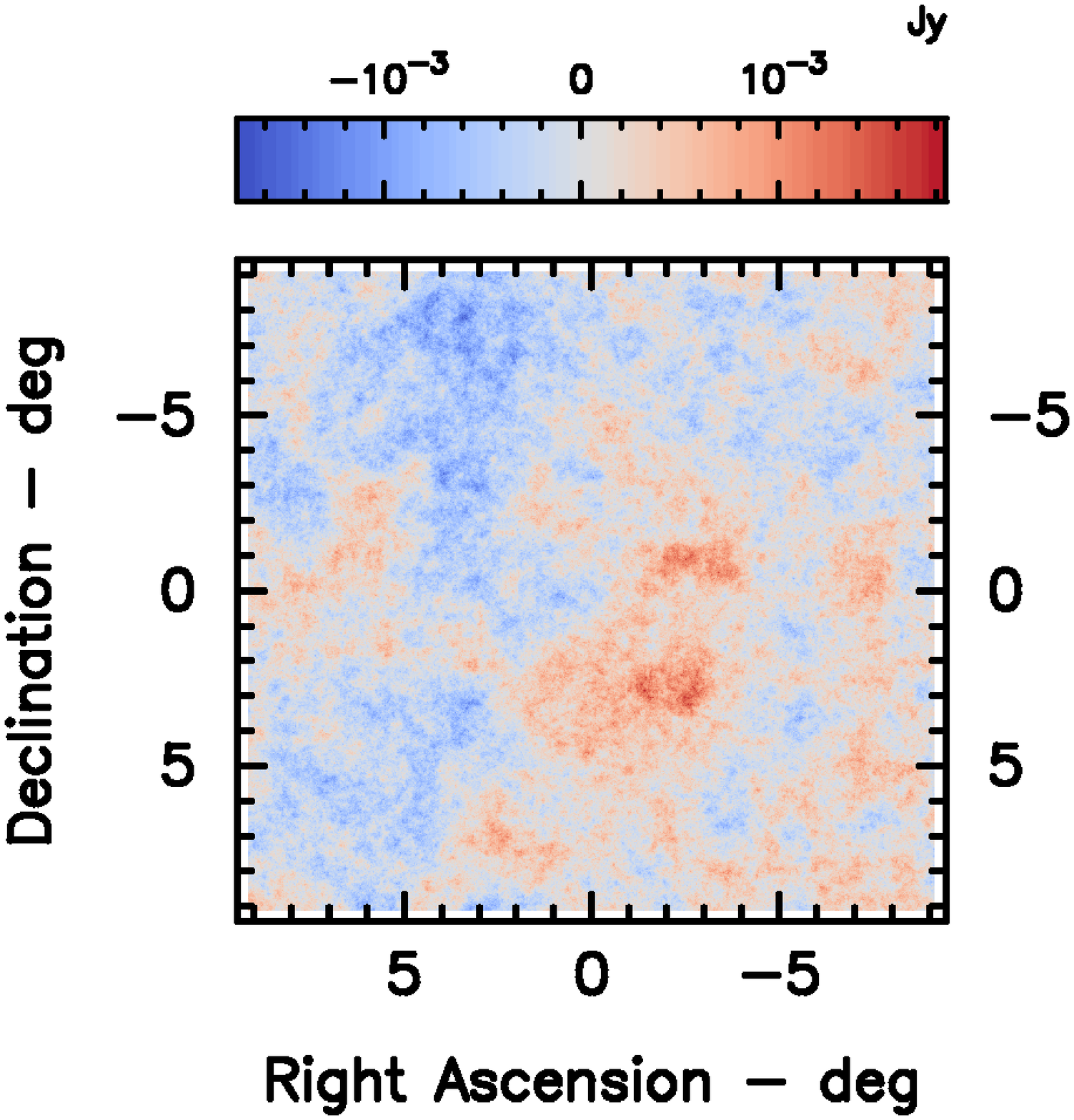}}
\hskip 3mm
\subfigure[Angular Power
  Spectrum]{\label{fig:inout-GAL}
\includegraphics[scale=0.28,angle=270]{powerspec.GAL.eps}} 
\hskip 7mm
\subfigure[Angular Power Spectrum with primary
  beam]{\label{fig:inout-GAL-pbeam}
\includegraphics[scale=0.28,angle=270]{powerspec.GAL.pbeam.eps}}
\end{minipage}
\caption{(colour online) (a) shows a particular realization of the simulated
brightness temperature map $\delta T({\btheta},\nu)$  in
$(\alpha,\delta)$ coordinates for the nominal frequency of 
$\nu=326.5 \, {\rm  MHz}$.In (b) the black points show the angular power
spectrum recovered  from the simulated map shown in (a),  the input model
angular power spectrum $C^M_{\ell}(\nu)$, is shown by the curve in red. The
shaded region shows the multipole range  $\ell_{\mathrm{min}}  
- \ell_{\mathrm{max}}$ accessible to OWFA. 
(c) shows the angular power spectrum recovered by applying 
equation~(\ref{eqn:ran111})  to the image obtained for a single realisation by multiplying the
simulated map (a)  with the primary beam patter shown in Figure
\ref{subfig:pbeam}. }
\end{figure*}

We assume that the fluctuations in
the diffuse Galactic Synchrotron radiation are statistically
homogeneous and isotropic, described by a Gaussian random field whose statistical
properties are completely specified by the angular power spectrum. We
further assume that the input model 
angular power spectrum $C^M_{\ell}(\nu)$ is well described by a
single power law in the entire range of angular scales of our
interest \citep[e.g.][]{Santos2005, Ali2008}:
\begin{equation}
C^M_{\ell}(\nu) = A_{\nu_0}
\left(\frac{\nu_0}{\nu}\right)^{2\alpha}\left(\frac{1000}{\ell}\right)^{\gamma}
\label{eqn:C_l}
\end{equation}
where $A_{\nu_0}$ is a constant - the amplitude that can be determined
observationally at $\nu_0$ (here at $\ell = 1000$), $\alpha$ is the
spectral index and $\gamma$ is the index of the spatial power. 

\begin{figure*}
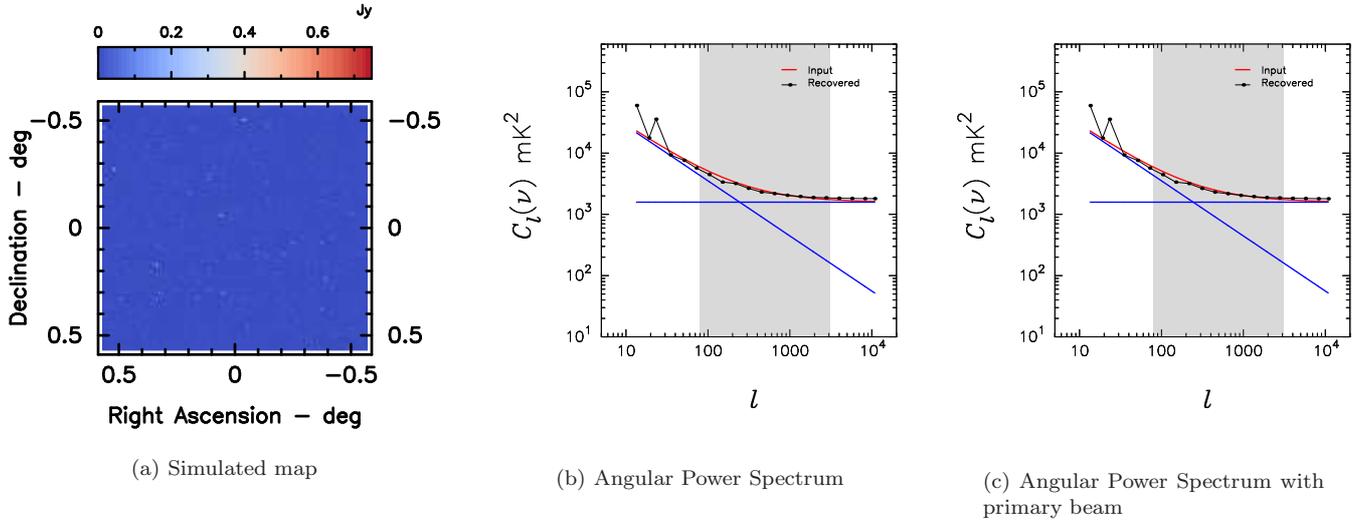

\centering
\begin{minipage}{190mm}
\subfigure[Simulated map]{\label{fig:EPS-map}\includegraphics[scale=0.235,angle=270]{EPSmap.eps}}
\hskip 7mm
\subfigure[Angular Power
  Spectrum]{\label{fig:inout-EPS}\includegraphics[scale=0.28,angle=270]{powerspec.EPS.eps}}
\hskip 6.7mm
\subfigure[Angular Power Spectrum with primary
  beam]{\label{fig:inout-EPS-pbeam}\includegraphics[scale=0.28,angle=270]{powerspec.EPS.pbeam.eps}}
\end{minipage}
\caption{(colour online) (a) shows a zoomed in view of a single realization of the simulated
brightness temperature map $\delta T({\btheta},\nu)$  in
$(\alpha,\delta)$ coordinates for the nominal frequency of 
$\nu=326.5 \, {\rm  MHz}$. In (b) the black points show the angular power
spectrum  recovered from the simulated map shown in (a),
the input model angular power spectrum $C^M_{\ell}(\nu)$, is shown by the
curve in red. The shaded region shows the multipole range $\ell_{\mathrm{min}}
- \ell_{\mathrm{max}}$ accessible to OWFA. (c) shows the angular power spectrum recovered by applying 
equation~(\ref{eqn:ran111})  to the image obtained for a single realisation by multiplying the
simulated map (a)  with the primary beam pattern shown in Figure
\ref{subfig:pbeam}.The approximation is valid throughout the entire $\ell$ range, except at
the smallest multipoles.}
\end{figure*}

A map of the diffuse synchrotron radiation can
be obtained as a realization of a Gaussian random field whose power is
distributed in the angular scales described by eqn~(\ref{eqn:C_l}).
A detailed description of how to generate such a realization can be found in
\citet{Choudhuri2014}, but 
we give a brief summary here for completeness. Consider the model angular
power spectrum given in equation~(\ref{eqn:C_l}). We take $A_{\rm 150} = 513\ 
{\rm mK}^2$, which is the amplitude of the angular power spectrum measured
at 150 MHz with the GMRT \citep{Ghosh2012}. For the other paramters, we take $\alpha
= 2.52$ \citep{Rogers2008} and $\gamma=2.34$ \citep{Ghosh2012}.
The spectral index $\alpha$ scales the observed brightness
temperature from a 150-MHz observation to 326.5 MHz as well as imparts
an in-band spectral shape to the diffuse emission. 

The foreground simulations have been carried out under the flat sky
approximation with $\mathbfit{n}=\mathbfit{m} + {\btheta}$, where $\btheta$
is a 2D vector on the plane of the sky. We use
$\mathbfit{U}$  to denote the Fourier modes corresponding to $\btheta$, and
these are related to the angular multipole $\ell$ as $\ell= 2 \pi |\!\mathbfit{U}\!|$.  The  Fourier components of 
the brightness temperature fluctuations are generated on a grid through
\begin{equation}
\Delta \tilde{T}(\mathbfit{U},\nu)=\sqrt{\frac{\Omega \,
    C^M_{\ell}(\nu)}{2}}\left(x+iy\right), 
\label{eqn:ran110}
\end{equation}
where $\Omega$ is the total solid angle of the simulated field, and 
$x$ and $y$ are independent Gaussian random variables with 
zero mean and unit variance. The map of the brightness temperature
fluctuations $\delta T({\btheta},\nu)$, or 
equivalently the specific intensity fluctuations 
$\delta I(\mathbfit{n},\nu)$, is obtained through a Fourier inversion.
Figure~\ref{fig:GAL-map} shows a single realization of the simulated
brightness temperature map in $\delta T(\btheta,\nu)$  in
$(\alpha,\delta)$ coordinates, for which the recovered angular power spectrum
$C_{\ell}(\nu)$ is shown in Figure \ref{fig:inout-GAL} in comparison with the
input model $C^M_{\ell}(\nu)$ used to obtain the random realization.
The simulated map is
multiplied with the OWFA primary beam pattern (Figure \ref{subfig:pbeam})
to obtain $\delta T_B({\btheta},\nu)=\delta
T(\btheta,\nu)\ A({\btheta},\nu)$ which is used to 
calculate the model visibilities (equation \ref{eqn:modelvis}). Here 
$\Delta \tilde{T}_B(\mathbfit{U})$, which is  the Fourier 
transform of  $\delta T_B({\btheta},\nu)$, is related to the angular power 
spectrum $C_{\ell}(\nu)$ as 
\begin{equation}
\langle |\Delta \tilde{T}_B(\mathbfit{U})|^2 
\rangle=\int d^2 \mathbfit{U}' |\tilde{A}(\mathbfit{U}-\mathbfit{U}')|^2
 C_{ \ell^{'}}(\nu) \
\label{eqn:ran112}
\end{equation}
where $\ell^{'}=2 \pi | \mathbfit{U}' |$ and  $\tilde{A}(\mathbfit{U})$ is
  the Fourier transform of the primary beam 
$A({\btheta},\nu)$.
Under the
assumption that $C_{\ell}(\nu)$ does not change much
within the width of the function $|\tilde{A}(\mathbfit{U}-\mathbfit{U}')|^2$, 
the relation between $\Delta \tilde{T}_B(\mathbfit{U})$ and 
the angular power spectrum $C_{\ell}(\nu)$ can be \citep{Ali2014} 
approximated as 
\begin{equation}
\langle | \Delta \tilde{T}_B(\mathbfit{U}) |^2  \rangle
=\left[\int d^2 \mathbfit{U}' |\tilde{A}(\mathbfit{U}')|^2
  \right] C_{\ell}(\nu) \,
\label{eqn:ran111}
\end{equation}
 which allows us also to estimate $C_{\ell}(\nu)$ from 
$\Delta \tilde{T}_B(\mathbfit{U},\nu)$ or, equivalently, from the model
visibilities $\mathbfit{M}(\mathbfit{U})$ .  Figure~\ref{fig:inout-GAL-pbeam}
shows the comparison between the  angular
power spectrum $C_{\ell}(\nu)$ estimated using equation~(\ref{eqn:ran111})  and
the input model $C^M_{\ell}(\nu)$. The recovered angular
power spectrum is in good agreement with the input model for $\ell \ge 100$,
whereas the convolution with the primary beam
(equation~\ref{eqn:ran112}) is important at smaller $\ell$. The fact
that $| \mathbfit{M}(\mathbfit{U}) |^2 $  directly gives an estimate of
$C_{\ell}(\nu)$ for most of the $\ell$ range shown in
Figure~\ref{fig:inout-GAL-pbeam} means that the correlation between the
visibilities measured at OWFA can be used directly to 
estimate the angular power spectrum of the sky signal, an issue that we
explore in detail later. 

The values of the three parameters $\alpha$, $A_{150}$ and $\gamma$
  have been held constant in our 
simulations.  In reality the spectral index $\alpha$ can vary along different lines of sight, and 
the amplitude $A_{150}$ and $\gamma$ is the index of the spatial power can have different values 
in different patches of the sky (e.g. \citealt{LaPorta2008}).  These variations will introduce 
additional angular and frequency structures as compared to the predictions of
our simulations.

\subsection{Extragalactic radio sources}
The dominant emission from the 326.5-MHz sky at most of the angular scales of
our interest is expected to be from the extragalactc point sources \citep{Ali2014}.
\citet{Wieringa1991} has measured the source counts at 325 MHz with the WSRT
down to $\sim 4$ mJy. More recently, \citet{Sirothia2009} have measured
 the source counts in the ELAIS N1 field at 325 MHz with the GMRT, the
 deepest till date. \citet{Ali2014} have estimated that at most of
 the angular scales of P-II and at all scales of P-I, the
 extragalactic point source contribution is likely to dominate when
 observing at high galactic latitudes. The extragalactic radio source power
 spectrum has a flat Poisson part and a clustered part with a non-zero slope
 \citep{Cress1996, Condon2007, Owen2008, Vernstrom2015}.
 
Simulating extragalactic radio sources, most of which are
unresolved at the resolution of OWFA, is more involved than the straightforward
Fourier inversion of a power spectrum of the diffuse foreground. 
Details of the method used here can be found in \citet{Gonzalez2005} but
we summarize it here for completeness. Let $\bar{n}$ be the mean number
density of sources per  pixel. Note that we have $\bar{n} \geq 1$ for a confusion
limited map. We begin with a random  Poisson field of mean source density
$\bar{n}$, for which the density  contrast is defined as
$\delta(\btheta) = [n(\btheta) -   \bar{n}]/\bar{n}$, and its Fourier
  transform is denoted using  
$\Delta(\mathbfit{U})$. 
Let its angular power spectrum of $\delta(\btheta)$
be denoted by $C_\ell^{P}$, which can be calculated from
$\Delta(\mathbfit{U})$. 
We modify $C_\ell^{P}$ by the clustering angular power spectrum  $C_\ell^{cl}$:
accordingly the Fourier transform of the density contrast is modified as
\begin{equation}
\Delta'(\mathbfit{U}) = \Delta(\mathbfit{U})\sqrt{\frac{{C_\ell^{P}+C_\ell^{cl}}}{{C_\ell^{P}}}}
\end{equation}
The modified density contrast spectrum $\Delta'(\mathbfit{U})$ is
reverse transformed to give the modified density contrast function
$\delta'(\btheta)$.
Finally, the modified pixel source density is given by $n'(\btheta) =
\tilde{n}[1 + \delta'(\btheta)]$. The input power spectra for the Poisson
part and the clustered part are respectively(from \citealp{Ali2014}),
\begin{equation}
C_\ell^{P} = \left(\frac{\partial B}{\partial T}\right)^{-2} \int_0^{S_c} S^2 \frac{dN}{dS} dS
\end{equation}
and
\begin{equation}
C_\ell^{cl} =  3.3 \times 10^{-5} \cdot \left(\frac{\partial B}{\partial T}\right)^{-2} \left[
  \int_0^{S_c} S^2 \frac{dN}{dS} dS\right] \left(\frac{1000}{\ell}\right)^{\gamma}
\end{equation}
The integrals have been computed from a polynomial fit to the source counts from 
\citet{Wieringa1991}. The simulated map represents the true source distribution
in the sky as no bright sources are assumed to have been subtracted, equivalent to an infinite
cutoff flux. However, we impose a cutoff of 30 Jy in the numerical integration:
this is a reasonably good approximation to an infinite cutoff flux, as the number of sources beyond $\sim
3$ Jy falls very sharply, and this behaviour is well captured by the polynomial
approximation to the differential source counts.
The clustered part has a power law index of $\gamma =
0.9$ \citep{Cress1996}. The mean values are $C_{1000}^{P} = 1580\ \mathrm{mK}^2$ and $C_{1000}^{cl}
= 444\ \mathrm{mK}^2$. We now populate the pixels with radio sources according to the observed
differential source count relation, drawing from the distribution described by
the polynomial obtained for the fit. The spectral indices
for the point source fluxes have been assigned randomly, drawing from
a Gaussian fit to the distribution centered at $\alpha = 0.7$ with a dispersion
of $\sigma_\alpha = 0.25$ \citep{Sirothia2009} and their
channel-wise fluxes scaled accordingly. 
The angular resolution of these simulated maps ($\sim 1\arcmin \times
1\arcmin$) is equal to the 325-MHz WSRT 
resolution of \citet{Wieringa1991}, much finer than the OWFA resolution of
$6\arcmin$. Figure~\ref{fig:EPS-map} shows a single realization of the simulated
brightness temperature map in $\delta T(\btheta,\nu)$  in
$(\alpha,\delta)$ coordinates. 
The angular power spectrum of the single random realization
$C_{\ell}(\nu)$, shown in Figure \ref{fig:inout-EPS}, is seen to be in good agreement with the input model
$C^M_{\ell}(\nu)$ except at
the smallest multipoles. Figure~\ref{fig:inout-EPS-pbeam}
shows the comparison between the angular
power spectrum $C_{\ell}(\nu)$ estimated using equation~(\ref{eqn:ran111})  and
the input model $C^M_{\ell}(\nu)$. 

The confusion limit for OWFA is 175
mJy \citep{Ali2014}; however, we do not propose to identify and
subtract sources from the one-dimensional images made using the OWFA
itself. Instead, a model for the sky obtained from deep GMRT imaging
could, for example, be used to subtract the contribution to the
visibility from the extragalactic radio sources. Alternatively, the
model visibilties upto 18 mJy (the $5\sigma$ lmit) could just as well
be obtained from the shallower Westerbork Northern Sky Survey (WENSS;
\citealt{Rengelink1997}).

\section{Power spectrum estimation}\label{sec:MAPS}
\subsection{The two-visibility correlation as the MAPS estimator}\label{ssec:V2MAPS}
The issue here is to use the measured OWFA visibilities
$\mathcal{V}(\mathbfit{U}_n)$  to estimate the statistics of the
background sky signal. Before discussing the method of analysis, it is
important to note that the baseline $\mathbfit{U}_n$ corresponding to a fixed
antenna separation $\mathbfit{d}_n$ scales as $\mathbfit{U}_n = \nu \,
\mathbfit{d}_n/c $  (equation \ref{eq:basln}) with $\nu$ varying 
across the observing band. This renders it difficult to use
$\mathbfit{U}_n$ and $\nu_i$ as independent parameters to label the visibilities 
$\mathcal{V}(\mathbfit{U}_n)$  that  will be measured at OWFA. For the
purpose of analyzing the data we  adopt the notation where  
$\mathbfit{U}_n = \nu_c \, \mathbfit{d}_n/c $ which does not vary with frequency
but is fixed at $\nu_c=326.5 \, {\rm  MHz}$.  In this notation we may interpret
$\mathbfit{U}_n$ and $\nu_i$ as  
independent parameters that label the visibilities
$\mathcal{V}(\mathbfit{U}_n, \nu_i)$. 

The proposed method of analysis is here applied to 
simulated data which only contains the foreground components. As described in
the previous section, the foregrounds all have simple power law frequency
dependence in the models  which we have implemented here. The telescope's
chromatic response   (equations \ref{eq:pb} and \ref{eq:basln}) , however,
introduces additional frequency dependence which we have attempted to capture
in the subsequent analysis. 

The two point statistics or, equivalently the power spectrum of the sky
signal, is entirely contained in the two visibility correlation 
\begin{equation}
\mathbfss{S}_2(\mathbfit{U}_n,\nu_i;\mathbfit{U}_m,\nu_j) \equiv 
\langle  \mathcal{V}(\mathbfit{U}_n,\nu_i) \mathcal{V}^{*}(\mathbfit{U}_m,\nu_j)
\rangle \,. 
\label{eq:s2a}
\end{equation}
where the angular brackets denote an ensemble average over different random
realisations of the sky signal. 

In addition to the self-correlation $(n=m)$ , 
the signal in the adjacent baseline pairs $(n=m \pm 1)$ are correlated
\citep[see Figure ~\ref{subfig:baseline-space} and ][]{Ali2014}. Further, the correlation is approximately a factor of four
smaller compared to the self correlation if we consider two adjacent
baselines \citep{Bharadwaj2015}. For the present analysis
we consider only the self-correlation, and we use the notation 
\begin{equation}
\mathbfss{S}_2(\mathbfit{U}_n,\nu_i,,\nu_j) \equiv 
\langle  \mathcal{V}(\mathbfit{U}_n,\nu_i) \mathcal{V}^{*}(\mathbfit{U}_n,\nu_j)
\rangle \,. 
\label{eq:s2b}
\end{equation}
 The relation between $\mathbfss{S}_2$  and the power spectrum
 $P(\mathbfit{k})$ of \ion{H}{I} is well understood
 \citep{Bharadwaj2001a,Bharadwaj2005,Ali2014} and we do not consider it here.

The multi-frequency angular power spectrum $C_\ell(\nu_i,\nu_j)$  
(MAPS; \citealt{Datta2007}) which is  defined through  
\begin{equation}
\langle \Delta \tilde{T}(\mathbfit{U},\nu_i)  \Delta
\tilde{T}_B(\mathbfit{U}^{'},\nu_j) \rangle =
\delta^2_D(\mathbfit{U}-\mathbfit{U}^{'}) C_{\ell}(\nu_i,\nu_j)
\label{eq:maps1}
\end{equation} 
provides an useful tool to jointly characterize
the angular and frequency dependence of the statistical properties of the sky
signal. Here $\delta^2_D(\mathbfit{U}-\mathbfit{U}^{'})$ is the 2D Dirac delta function
and we have $\ell=2 \pi |\mathbf{U}|$  
in the flat sky approximation.  

The model visibilities in equation \ref{eqn:model_RIME} are related to the
brightness temperature fluctuations $\Delta \tilde{T}(\mathbfit{U},\nu)$
as
\begin{equation}
\mathbfit{M}\left(\mathbfit{U}_{n},\nu \right)=
\left(\frac{\partial B}{\partial  T}\right)_{\nu}
\int d^2 \mathbfit{U} \, \tilde{A}(\mathbfit{U}_{n} (\nu/\nu_c)-\mathbfit{U} ,\nu)
\Delta \tilde{T}(\mathbfit{U},\nu)
\label{eq:maps2}
\end{equation}
where $\left(\frac{\partial B}{\partial  T}\right)_{\nu}$ is the conversion
factor from brightness temperature to specific intensity, and we have
introduced a  factor $(\nu/\nu_c)$ in the argument of $\tilde{A}$ to account for
the fact that  
$\mathbfit{U}_{n} $ is defined at $\nu_c$. For the present
analysis we assume 
that the visibilities are perfectly calibrated and altogether ignore the noise contribution. 
This leads to 
\begin{eqnarray}
&& \mathbfss{S}_2(\mathbfit{U}_n,\nu_i,,\nu_j)=
 \left(\frac{\partial B}{\partial  T}\right)_{\nu_i}
\left(\frac{\partial B}{\partial  T}\right)_{\nu_j} 
\int d^2 U \, \times 
\label{eqn:maps3}
 \\
&&  \tilde{A}(\mathbfit{U}_{n} (\nu_i/\nu_c)-\mathbfit{U} ,\nu_i) 
\tilde{A}^{*}(\mathbfit{U}_{n} (\nu_j/\nu_c)-\mathbfit{U} ,\nu_j) 
C_{2 \pi U}(\nu_i,\nu_j) \nonumber
\end{eqnarray}
which relates the visibility correlation to the multi-frequency angular power
spectrum (MAPS). Note that it is possible to estimate the  visibility
correlation $\mathbfss{S}_2(\mathbfit{U}_n,\nu_i,,\nu_j)$  directly from the
measured visibilities,  whereas  $C_{\ell}(\nu_i,\nu_j)$ quantifies the
intrinsic sky signal independent of any telescope. Equation (\ref{eqn:maps3}) allows us to
relate the statistics of the measured visibilities to the intrinsic statistics
of the sky signal. We see that the telescope appears 
through a convolution in equation (\ref{eqn:maps3}). It is, in principle,
possible to determine $C_{\ell}(\nu_i,\nu_j)$ from the measured 
$\mathbfss{S}_2(\mathbfit{U}_n,\nu_i,,\nu_j)$ by 
deconvolving the  effect of the telescope's primary beam $\tilde{A}$, and we 
interchangeably refer to $\mathbfss{S}_2(\mathbfit{U}_n,\nu_i,,\nu_j)$ as the
visibility correlation or the MAPS estimator. 

We now consider the restricted situation where $\nu_i=\nu_j$, and discuss how
it is possible to use the measured $\mathbfss{S}_2(\mathbfit{U}_n,\nu_i,\nu_j)$ to
estimate $C_{\ell}(\nu) \equiv C_{\ell}(\nu_i,\nu_i)$. We have already seen in equation
(\ref{eqn:ran110}) that for a large part of the $\ell$ range probed by OWFA 
it is possible to approximate the convolution in
equation (\ref{eqn:maps3}) by a multiplicative factor . This permits us to
approximate $\mathbfss{S}_2(\mathbfit{U}_n,\nu_i,\nu_j)$ as  
\begin{equation}
\mathbfss{S}_2(\mathbfit{U}_n,\nu,\nu) = K_{\nu}
C_{\ell_n}(\nu)
\label{eq:maps4}
\end{equation}
where $\ell_n=2 \pi |\mathbfit{U}_n|$ and $K_{\nu}$ is defined as  
\begin{equation}
K_{\nu}=\left(\frac{\partial B}{\partial  T}\right)_{\nu}^2 
\int d^2\mathbfit{U}' |\tilde{A}(\mathbf{U}')|^2  \,,
\label{eq:maps5}
\end{equation}
which is uniquely determined for a known primary beam.
We can use equation (\ref{eq:maps4}) to estimate
$C_{\ell}(\nu)$ from the measured visibility correlation. 
 
Considering a fixed baseline $\mathbfit{U}_n$, the visibility correlation 
$\mathbfss{S}_2(\mathbfit{U}_n,\nu_i,\nu_j)$ is (equation \ref{eqn:maps3})
a real valued symmetric matrix of dimensions $N_c \times N_c$ where
$N_c$ is the number of frequency channels. Observations will yield
such a matrix for every available unique baseline $\mathbfit{U}_n$. For OWFA 
(Table~\ref{tab:OWFA}) we have a limited number of such baselines all
of which are aligned with the N-S direction, allowing us to label the
baselines  with just numbers $U_n$ instead of vectors $\mathbfit{U}_n$.
Earlier studies  \citep[e.g.][]{Bharadwaj2001a, Ali2014} show that the
\ion{H}{I} signal 
contribution to  $\mathbfss{S}_2({U}_n,\nu_i,\nu_j)$  can be modelled as 
$\mathbfss{S}_2({U}_n,\Delta \nu_{ij})$
  which depends only on  the
frequency separation $\Delta \nu_{ij}=| \nu_i-\nu_j |$.  It is further
predicted that $\mathbfss{S}_2({U}_n,\Delta \nu_{ij})$  has a maximum value
at $\Delta \nu_{ij}=0$, and decorrelates rapidly as the separation $\Delta \nu_{ij}$ increases, with a value close to zero 
for large values of $\Delta \nu_{ij}$ \citep[see e.g.][for result from simulation]{Chatterjee2017}. $\mathbfss{S}_2({U}_n,\Delta \nu)$  is
predicted to be close to zero for $\Delta \nu_{ij} \ge 1 \, {\rm MHz}$ on nearly all
of the baselines that will be probed by P-I of OWFA \citep{Ali2014}. This
essentially implies that the \ion{H}{I} signal is mainly localized near the diagonal
elements of the visibility correlation matrix
$\mathbfss{S}_2(\mathbfit{U}_n,\nu_i,\nu_j)$, the elements which are located far 
away from the diagonal are expected to not have much of the \ion{H}{I} signal.  It
is therefore reasonably safe to assume that the matrix elements located at a
distance from the diagonal are entirely dominated by the foregrounds with a
negligible \ion{H}{I}  contribution.
%The idea is to use these matrix elements to
%model the $\nu_i$ and $\nu_j$ dependence of the foreground component of 
%the measured $\mathbf{S}_2(\mathbf{U}_n,\nu_i,,\nu_j)$, and then extrapolate
%this model to the diagonal and near diagonal elemnts where it can be
%subtracted out to recover the \ion{H}{I} signal. The full implementation and
%validation of this proposed foreground removal technique is beyond the scope
%of this paper. The proposed technique relies on the assumption that the
%foreground component of  the measured $\mathbf{S}_2(\mathbf{U}_n,\nu_i,,\nu_j)$ 
%has a slow, smooth  $\nu_i$ and $\nu_j$ dependence which is crucial for the
%extrapolation to work. While the foregrounds are expected to have a smooth 
%intrinsic frequency dependence, the chromatic telescope response  could possibly
%introduce frequency dependent structures which may pose a challenge for the
%proposed scheme. 

In the current work we present preliminary results 
where we explore the frequency dependence of
$\mathbfss{S}_2(\mathbfit{U}_n,\nu_i,\nu_j)$ in order to assess the
contribution to such effects from the instrument itself. We note that 
an earlier version of the visibility correlation-based estimator presented here has 
been applied  to  measure  the foreground  $C_{\ell}$ in $150 \, {\rm MHz}$
GMRT observations \citep{Ali2008,Ghosh2012},  and also for foreground removal
in $610 \, {\rm MHz}$ GMRT observations \citep{Ghosh2011a,Ghosh2011b}. 

\begin{figure}
\centering
\includegraphics[scale=0.35, angle=270]{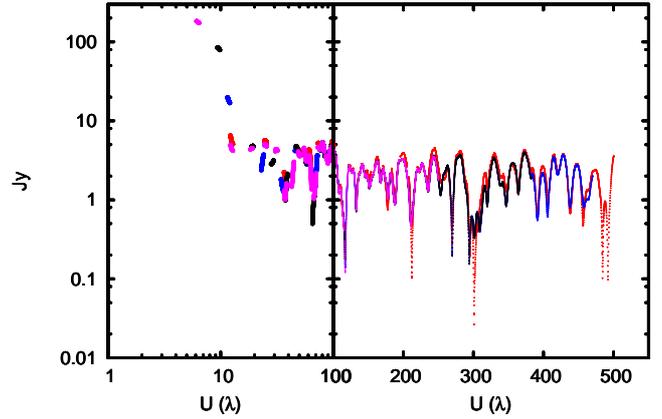}
\caption{(colour online) The red, blue, black and magenta colours code
  respectively the declinations $0^\circ$, $20^\circ$, $40^\circ$ and
  $60^\circ$. This figure shows the magnitude of the model
  visibilities for a   single realization of the sky, the emission from
  the larger scales being   picked up by the shortened baselines at increasing 
  declination.}
\label{fig:s_vs_u}
\end{figure}
\subsection{Computing the estimator}\label{ssec:V2}
In practice, we would calibrate the observed visibilities making use
of the redundant spacings \citep[e.g.][]{Wieringa1992, Liu2010,
  Marthi2014}.  
In this simulation we directly use the model visibilities, equivalent 
to the observed visibilities having been already
calibrated. For a baseline $\mathbfit{U}_n$ and  a frequency channel 
$\nu_i$, let there be $N_n$ distinct measurements  of the visibility
$\mathcal{V}^{(a)}(\mathbfit{U}_n, \nu_i)$ labelled using $a=1,2,...,N_n$.
Note that each redundant spacing as well as each time sample is
a distinct measurement of the visibility\footnote{This is true only when the
  telescope tracks the sky continuously. We do not consider the drift-scan mode
of observing in this article.}. The $N_n$ distinct measurements all
measure the same Fourier mode on the sky, given by
the model visibilities $\mathcal{M}(\mathbfit{U}_n, \nu_i)$.
The system noise contribution $\mathcal{N}^{a}\left(\mathbfit{U}_{n} ,\nu \right)$ 
(equation \ref{eqn:RIME}), however, is different in each of these measurements:
its contribution to each visibility measurement is an independent
Gauusian random variable, and the different visibility measurements can be  
added coherently (after calibration) to improve the signal-to-noise ratio; this
is a unique feature of OWFA. Therefore, the sum 
\begin{equation}
\bar{\mathcal{V}}(\mathbfit{U}_n,\nu_i) = \frac{1}{N_n} \, \sum_{a=1}^{N_n}
\mathcal{V}^{(a)}(\mathbfit{U}_{n}, \nu_i) 
\end{equation}
shows a baseline-dependent $1/\sqrt{N_n}$ noise behaviour.

We can, in principle, use $\bar{\mathcal{V}}(\mathbfit{U}_n,\nu_i) 
\bar{\mathcal{V}}^{*}(\mathbfit{U}_n,\nu_j)$ to estimate the visibility
correlation $\mathbfss{S}_2(\mathbfit{U}_n, \nu_i, \nu_j)$. The correlation 
$\mathbfss{S}_2(\mathbfit{U}_n, \nu_i, \nu_i)$, 
however, will pick up an extra, positive, noise bias 
$\sum_{a=1}^{N_n} | \mathcal{N}^{a}\left(\mathbfit{U}_{n} ,\nu_i \right)
|^2/(N_n)^2$  arising from the correlation of a visibility with itself. 
It is possible to avoid this noise bias \citep[e.g.][]{Begum2006, Choudhuri2016b} by subtracting out
the self correlation 
\begin{equation}
\bar{\mathcal{V}}_2(\mathbfit{U}_n,\nu_i) = \frac{1}{N_n} 
 \sum_{a=1}^{N_n} |\mathcal{V}^{(a)}(\mathbfit{U}_{n}, \nu_i)|^2 \,.
\end{equation}
We hence define our estimator $\mathbfss{S}_2$ as
\begin{equation}
\mathbfss{S}_2(\mathbfit{U}_n, \nu_i, \nu_j) = \frac{N_n^2 \,
  \bar{\mathcal{V}}(\mathbfit{U}_n,   \nu_i)
\bar{\mathcal{V}}^*(\mathbfit{U}_n, \nu_j) -
  \delta_{ij} N_n \,  
\bar{\mathcal{V}}_2(\mathbfit{U}_n,   \nu_i)}{N_n^2 - \delta_{ij}  \, N_n}
\label{eqn:S_2_vs_nu1_nu2}
\end{equation}
which is free from the noise bias arising from the correlation of a measured
visibility with itself. 

\section{Results}\label{sec:results}
\subsection{Visibilities}\label{subsec:vis}
The model visibilities are computed
for a sky model that consists of the sum of a realization of the diffuse
foreground and the extragalactic
radio sources. The full
$\mathbfit{U}$-range of the OWFA is $\sim 0-500\lambda$
at $\delta_0 = 0^\circ$. However, for fields at high
declinations, the range of $\mathbfit{U}$ is compressed by the factor $\cos \delta_0$, where
$\delta_0$ is the declination of the centre of the
field. 
The model visibilities for a single realization of the
sky that includes both the diffuse emission and extragalactic radio
sources for four different declinations are shown in
Figure~\ref{fig:s_vs_u}. 
The higher declinations allow for a compressed range of
$\mathbfit{U}$, thereby accessing more of the extended emission at $\mathbfit{U}
< 100$. However, the sensitivity reduces equally by $\cos \delta_0$ due to the
smaller projected aperture. 

\subsection{The Foreground Power Spectrum}\label{subsec:MAPS_example}
\begin{figure*}
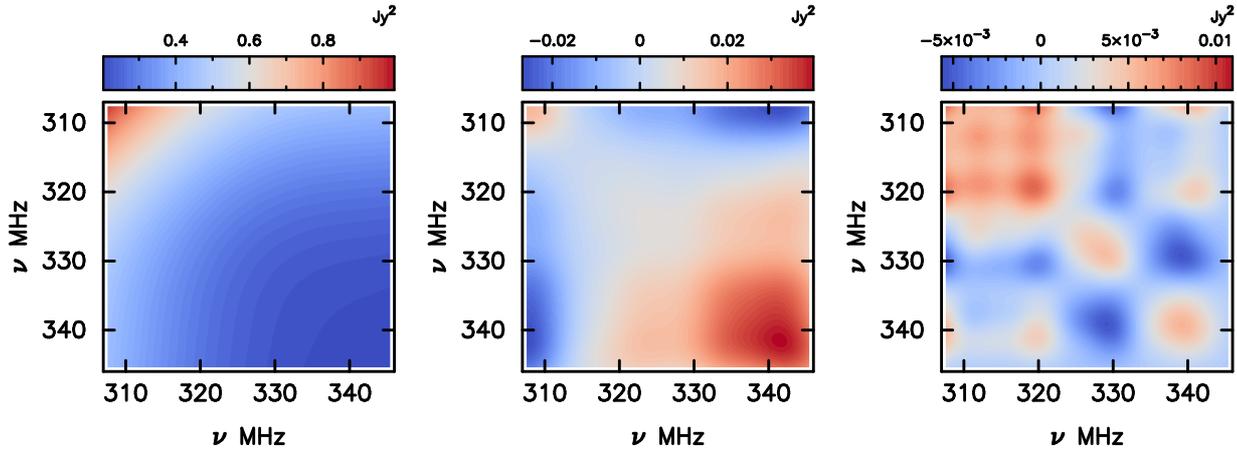

\begin{center}
\begin{minipage}{190mm}
\includegraphics[scale=0.25, angle=270]{color_V2_BL4.eps}
\hskip 2mm
\includegraphics[scale=0.25, angle=270]{color_V2_BL21.eps}
\hskip 2mm
\includegraphics[scale=0.25, angle=270]{color_V2_BL31.eps}
\end{minipage}
\caption{(colour online) The MAPS estimator $\mathbfss{S}_2(\mathbfit{U}_n,\nu_i,
  \nu_j)$ for a single realization of the sky, for simulated 
  observations at $\delta = 0^\circ$ for OWFA
  baselines $n=1, 16$ and $32$ increasing from left to right panel. The
    diagonal in each matrix is the locus of $\Delta\nu = 0$, and any point on
  the diagonal through the MAPS for the baselines stacked in sequence is a
  proxy for $C_\ell(\Delta\nu=0)$. In addition to the shorter baselines having
  more power, the longer baselines exhibit enhanced spectral structure, both in
  line with our understanding. See text for details}.
\label{fig:V2_MAPS}
\end{center}
\end{figure*}

In this section we  provide results for the visibility correlation 
MAPS estimator. As a verification step, we start with a noise-free
simulation, and use the model visibilities 
directly: the assumption of perfect calibration and absence of ionosphereic effects and RFI
is implicit. The MAPS estimator returns a matrix of the visibility correlation
$\mathbfss{S}_2(\mathbfit{U}_n, \nu_i, \nu_j)$ as a function of the pair of
frequencies at which it is computed. Each panel (left to right)  in
Figure~\ref{fig:V2_MAPS} is the matrix (read top left to bottom right)
$\mathbfss{S}_2(\mathbfit{U}_n, \nu_i, \nu_j)$ for the OWFA baselines $n=1, 16$ and
$32$ respectively.
 
This observation is simulated for a declination of $\delta = 0^\circ$ and
includes the diffuse Galactic emission as well as the point source contribution.
The three panels of Figure~\ref{fig:V2_MAPS} (left to right), have different
(decreasing) foreground amplitudes.
As discussed in Figure 9 of \citet{Ali2014}, the diffuse Galactic emission is the dominant 
component at shortest baseline as apparent in the left panel of the Figure~\ref{fig:V2_MAPS}.
The clustered point source component dominates at small angular scales or the
larger baselines $(\mathbfit{U} \lesssim 300)$ which can be seen in the middle
panel. The foreground amplitude right panel where $\mathbfit{U} \simeq 400$, is
dominated by the Poisson contribution.
%As mentioned earlier, the \ion{H}{I} signal (not included here) is expected
%to be localized in the vicinity of the diagonal (top left to bottom right) of
%each panel of Figure~\ref{fig:V2_MAPS}, and is expected to be negligible 
%when $| \nu_i - \nu_j  | \geq 1 \, {\rm MHz}$
The foregrounds appear to be quite smooth in the $\nu_i-\nu_j$ plane,
suggesting that this could possibly be used to separate the \ion{H}{I} signal, which
has a different signature in $\mathbfss{S}_2$, from the foregrounds.  
Note that the foregrounds exhibit a  particularly smooth frequency dependence
at the smallest baseline (left panel), and the frequency variations become
progressively more rapid as we move to larger baselines. Much of the
frequency structure seen in the central and right panels is clearly not a 
function of the frequency separation $\Delta \nu =| \nu_i - \nu_j  |$
alone, and these structures are visible even at large $\Delta \nu $,  
 or away from the diagonal. This distinction between the
 foregrounds and the \ion{H}{I} signal leads to 
   separable features between the foregrounds and \ion{H}{I} in the $k$-space.

\begin{figure*}
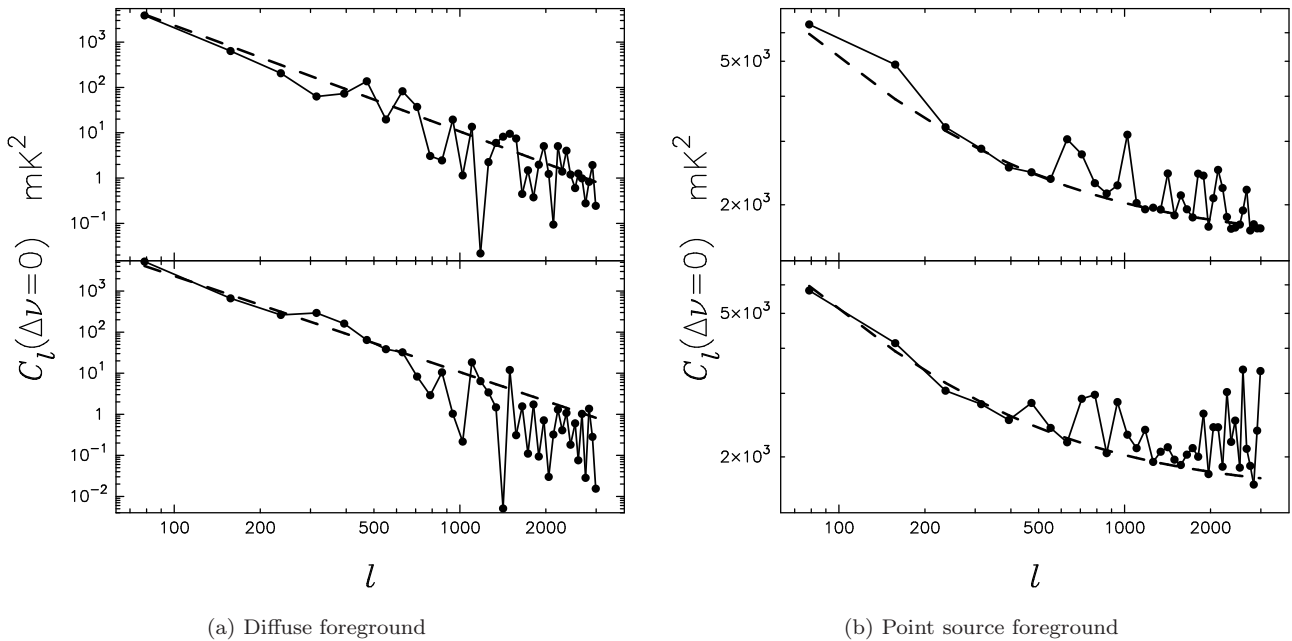

\centering
\begin{minipage}{190mm}
\subfigure[Diffuse
  foreground]{\label{subfig:GAL_PS}\includegraphics[scale=0.45,
    angle=270]{GSE-realizations-input-vs-recovery.eps}}
\hskip 5mm
\subfigure[Point source
  foreground]{\label{subfig:EPS_PS}\includegraphics[scale=0.45,
    angle=270]{EPS-realizations-input-vs-recovery.eps}}
\end{minipage}
\label{fig:PS}
\caption{(a) The angular power spectrum of the diffuse
  Galactic emission: input model shown with the solid line and the power
  spectrum recovered through the estimator with points, for two random realisations.
(b) The angular power spectrum of extra-galactic point sources.
In both the plots, the dashed line represents the analytical expression for the
power spectra used to obtain the random realisations, shown here only to compare
the random realisations against. The results are at the frequency $\nu_c$ = 326.5 MHz.} 
\end{figure*}

We next use the diagonal elements $\mathbfss{S}_2(\mathbfit{U}_n,\nu_i,  \nu_i)$
of the measured MAPS to estimate the angular power spectrum $C_{\ell_n}(\nu_i, \nu_j)$
of the sky signal using  equation (\ref{eq:maps4}). 
Figure~\ref{subfig:GAL_PS} shows $C_{\ell}(\nu)$
recovered from two simulated realisations of the diffuse Galactic
foregrounds. The solid line is the input analytical angular power spectrum and the
points represent the recovered 
angular power spectrum for a single realisation. Similarly,
Figure~\ref{subfig:EPS_PS} shows the recovered angular power spectrum for the
extragalactic point sources, where the solid line is the input
angular power spectrum, which is the sum of the Poisson and the clustered
contributions. We see that the recovered angular power
spectrum is in excellent agreement with the input model for the entire $\ell$
range shown in the figure. In both cases, the dashed line represents the model
used to obtain a random realization, and serve only as a visual aid.
We reiterate that the results are shown here for a
noiseless simulation. The MAPS estimator, in addition, is unbiased as it avoids
the self-noise contribution through equation (\ref{eqn:S_2_vs_nu1_nu2}).
In a future article, we will explore applying the tapered gridded estimator
(TGE; \citealt{Choudhuri2014}), tailoring it suitably for OWFA.

\section{Discussion}\label{sec:discussion}
We will now discuss some properties of the estimator in the context of the OWFA experiment, including limitations
posed by the instrument systematics. We will look at the
following aspects in succession:
\begin{itemize}
\item The estimator shows predictable spectral behaviour and introduces no
  spectral anomalies.
\item The chromatic primary beam and the chromatic response function of the
  interferometer dominate well over the chromaticity of the measured sky signal. 
\item Contribution from sources in the sidelobes manifest as spectral
  signatures in the estimator.
\item The error on the estimator can be derived from the
  data.

\end{itemize}
\subsection{Spectral behaviour}
The MAPS estimator has a fully tractable spectral response. Since it is
  possible to analytically derive the spectral response of the estimator, we can
  ascertain through simulations that the estimator introduces no unaccounted
  features in the spectrum. This means that, in principle, it is possible to
  model out the instrument-induced spectral features in the estimator.

It has been shown that the angular
power spectrum $C_\ell(\nu)$ and the MAPS estimator 
$\mathbfss{S}_2(\mathbfit{U}, \nu_i, \nu_j)$ are related by
equation (\ref{eqn:maps3}), where for our assumed sinc-squared primary beam, 
\begin{equation}
\tilde{A}(\mathbf{U},\nu) =
\frac{\lambda^2}{bd}\ \Lambda\!\left(\ \frac{u \lambda}{d}\ \right)
\Lambda\!\left(\ \frac{v \lambda}{b}\ \right)
\end{equation}
is the Fourier transform of the primary beam power pattern, 
$\mathbf{U} = (u, v)$ and $\Lambda\!(x)$ is the triangular
function defined as 
\begin{equation}
\Lambda\!(x) = 1 - |x|\ \mathrm{for}\ |x| <
1,\ \mathrm{and}\ \Lambda\!(x) = 0\ \mathrm{for}\ |x| \ge 1.
\end{equation}
Given that $\ell = 2 \pi U$ where $U = d/\lambda$, we may write 
\begin{equation}
C_\ell \propto \left(\frac{\lambda}{d}\right)^{\gamma}
\end{equation}
It then follows that for the visibility correlation of the diffuse
foreground signal alone, 
\begin{equation}
\mathbfss{S}_2 \propto \nu^{2(1-\alpha) - \gamma} 
\label{eqn:V2_behaviour}
\end{equation}
considering all the spectral contributions, namely the intrinsic
spectral index $\alpha$, the power law index dependence on frequency
through $\ell$, and the Fourier transform of the primary beam power
pattern. 
\begin{figure*}
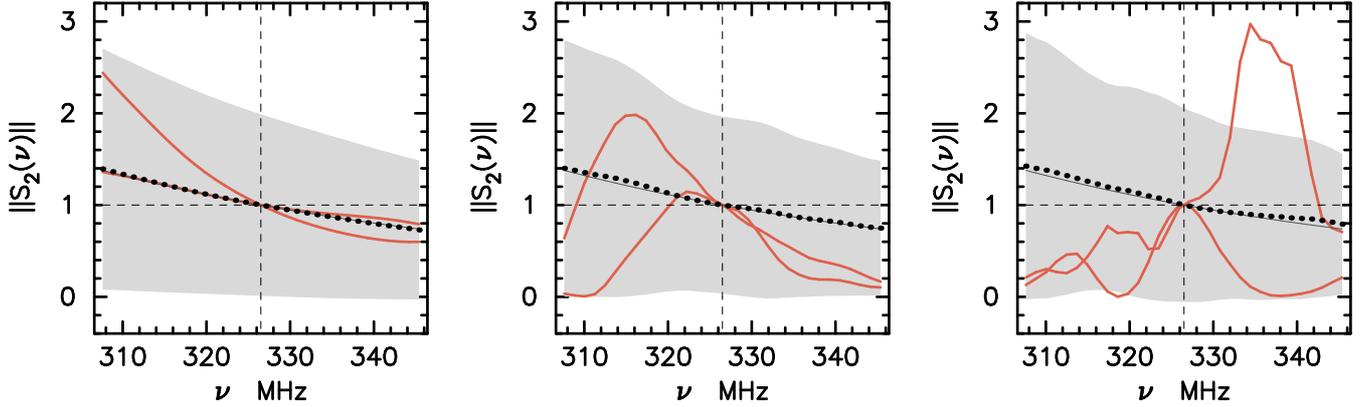

\begin{center}
\begin{minipage}{190mm}
\includegraphics[scale=0.28, angle=270]{BL4.v2-vs-nu.eps}
\hskip 4mm
\includegraphics[scale=0.28, angle=270]{BL21.v2-vs-nu.eps}
\hskip 4mm
\includegraphics[scale=0.28, angle=270]{BL31.v2-vs-nu.eps}
\end{minipage}
\caption{(colour online) The normalised MAPS estimator $\parallel\!\mathbfss{S}_2(\mathbfit{U},\nu)\!\parallel$ is
  shown left to right, top to bottom, for baselines 4, 21 and 31. The red curves relate to two different
  realisations of the diffuse Galactic synchrotron foreground from a total of
  1000. The estimated mean is given by the filled circles and the solid black line is the analytically
  computed curve given by $(\nu/\nu_0)^{(2-2\alpha-\gamma)}$. The simulation is
  over 39 MHz split into 312 channels, but only a sampled version of the
  estimated mean (filled circles) is shown for clarity. The shaded region
  represents the $1\sigma$ error bound on the random foreground realisations.}
\label{fig:V2-vs-nu}
\end{center}
\end{figure*}

Restating that the spectral response of the estimator is fully
  tractable, we illustrate this for a simple case where $\alpha = 2.32$
and $\gamma = 2.54$. The $\mathbfss{S}_2$ matrix is obtained for 1000 independent
but noiseless realisations of the diffuse Galactic foreground and the mean and the $1\sigma$
error bars of these diagonals are computed. 
The mean and the error bars are then individually normalised at $\nu = 326.5$ MHz; the normalised mean
and error bar curves can be represented as $\|\mathbfss{S}_2(\nu)\|$ and
$\|\Delta\mathbfss{S}_2(\nu)\|$. Figure~\ref{fig:V2-vs-nu} shows the normalised
mean of the diagonals as filled black circles. These have been sampled at every tenth
channel so that they are seen clearly in the plot. The shaded region represents
the bounds of the $1\sigma$ error on the mean of the foreground realisations. The
curves in red are the normalised (at
326.5 MHz) $\mathbfss{S}_2(\nu)$ curves for two different realisations. The
black curve is the mean of all the realisations and it
agrees remarkably well with the expression
$\left(\frac{\nu}{\nu_0}\right)^{2\left(1-\alpha\right)-\gamma}$.
We make a few points to enable a clear interpretation of
  Figure~\ref{fig:V2-vs-nu}. Firstly, the error bounds on the analytical curve
  arise from the unit variance complex random variable $(x + iy)/\sqrt{2}$ in equation
  (\ref{eqn:ran110}). Therefore, it shows the standard deviation of the random
  realisations and not the estimator. The standard deviation computed from the
  realisations is indeed unity,
  which is easy to see at $\nu_c$ = 326.5 MHz. In addition, the red curves show just
  two of such random realisations, and there is visibly more spectral structure
  on the longer baselines. As a result of which, lastly, the match is not
  visibly exact for the longer baseline between the analytical curve and the
  estimated mean data points. A progressively larger number of realisations would have to be
  averaged for the match to appear exact on the longer baselines. In addition,
  the shaded bounds allow for a comparison of these two realisations with the
  error on the mean, clearly showing more variance on the longer baselines.

A source away from the phase centre leaves oscillatory features in the
MAPS estimator, and the frequency of the oscillation depends on the location of
the source $m$ as well
on the baseline $nd$. It is interesting to note that
$\Delta\nu/\nu_0$ is of the order of 12\% ($\sim$40 MHz/327 MHz)
for OWFA.
\begin{figure}
\includegraphics[scale=0.5]{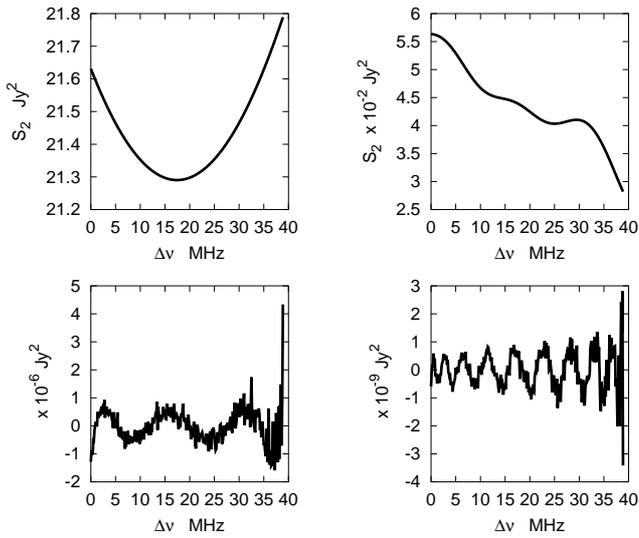}
\caption{The $\mathbfss{S}_2(\Delta\nu)$ curves for two baselines $\mathbfit{U}_2$
  and $\mathbfit{U}_{17}$ are shown in the top left and right panels
  respectively. The residual after fitting polynomials are shown in
  the bottom panels, indicating the contribution from sources in the sidelobes.}
\label{fig:v2_res}
\end{figure}
The argument can be generalised to include many
point sources, and by extension to the diffuse foregrounds as well. 
The sum total contribution from all emission within the field of view
superimpose with a range of phases that tend to partly 
cancel out. But residual features remain imprinted on the
innocuous-looking smooth spectra. Figure~\ref{fig:v2_res} shows the
$\mathbfss{S}_2$ for the same realisation of diffuse and point source
foregrounds used to obtain the plots in Figure~\ref{fig:V2_MAPS}, but computed
as a function of $\Delta\nu$ by averaging along
the diagonals of the $\mathbfss{S}_2$ matrix for two example baselines. Although
the foregrounds are non-stationary as discussed in Section~\ref{ssec:V2},
it is instructive to cast the estimator in the familiar form as a function of
$\Delta\nu$ for this exercise.
The apparently smooth curves in the top panels can be fitted by low-order polynomials successively.
The residual contamination can still be seen in the bottom panels. 
The frequency of the oscillatory pattern is higher as expected for longer
baselines. It must be noted that the amplitude of the residual oscillatory features is about 7
orders smaller than the visibility correlation $\mathbfss{S}_2$. These features
are still about 1-2 orders above the amplitude of the expected \ion{H}{I}
visibility correlation \citep[see][]{Ali2014, Chatterjee2017}. At this level,
these dominant residual features preclude detecting the \ion{H}{I} signal in the
$\Delta\nu$ space.
\begin{figure*}
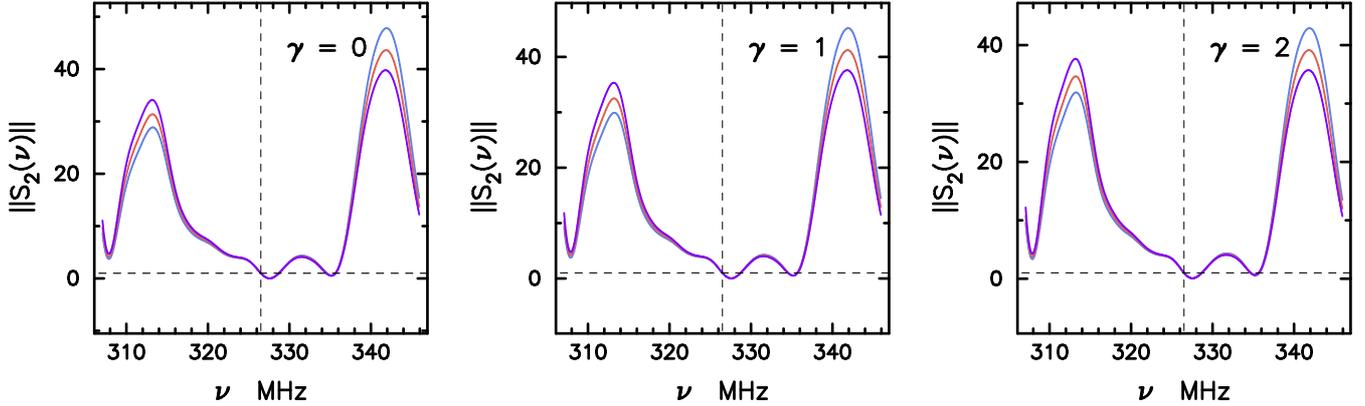

\centering
\begin{minipage}{190mm}
\includegraphics[scale=0.28, angle=270]{V2_BL38_BETA_0.eps}
\hskip 4mm
\subfigure{\includegraphics[scale=0.28, angle=270]{V2_BL38_BETA_1.eps}}
\hskip 4mm
\includegraphics[scale=0.28, angle=270]{V2_BL38_BETA_2.eps}
\end{minipage}
\caption{(colour online) The estimator $\mathbfss{S}_2(\nu)$ for different values of
$\alpha$ and $\gamma$ are shown here for the longest baseline $|\mathbfit{U}| = 475.0$. Each plot has three curves; one each for
$\alpha = 0.0, 1.0, 2.0$, shown by the colours blue, red and
purple respectively. The values have been normalised to
$\mathbfss{S}_2(\nu)$ at $\nu = 326.5$ MHz. The normlized curves look very similar
between widely different values for $\alpha$ and $\gamma$, implying that these
estimates are dominated not by $\alpha$ and $\gamma$, but by the instrument
response. The dashed lines mark $\nu_c$ = 326.5 MHz and $|\!|S_2(\nu_c)|\!| = 1$.}
\label{fig:V2_vs_beta}
\end{figure*}

These undesirable oscillatory features can be suppressed by
restricting the total FoV. This can be achieved
by tapering the primary beam with a weighting function, typically a
Gaussian window function. However, this is possible only in the $uv$-plane
post-correlation, through a 2D convolution. Beam tapering in $uv$ has been shown
to be quite effective for the GMRT \citep{Ghosh2011b}. The tapered-gridded
estimator has been shown to be equally effective for 
simulated GMRT data \citep{Choudhuri2014, Choudhuri2016a}.

The function $\mathbfss{S}_2(\mathbf{U}, \nu_1 - \nu_2)$
behaves smoothly away from
and normal to the principal diagonal. Since this cut normal to the diagonal
represents $\mathbfss{S}_2(\mathbf{U}, \Delta\nu)$, the MAPS estimator
  matrix $\mathbfss{S}_2$ naturally encodes the decorrelation of the foreground
  as well as the sky signal.

\subsection{Sources in the sidelobes and instrument chromaticity}\label{subsec:sidelobe}
The effects of sources at large angular distances from the pointing
centre coupling in through the primary beam is
well-known \citep[see e.g.][]{Datta2010, Vedantham2012,
  Pober2013a, Thyagarajan2015a, Thyagarajan2015b, Pober2016}. Consider a single point
source at the co-ordinates ${\boldsymbol \theta} = (l, m)$. Let us also assume that the
pointing of observation is $(\alpha_0 = 0, \delta_0 = 0)$ without loss of
generality. 
For the $n^{th}$ baseline  in the linear array, we note that at frequency $\nu=c/\lambda$, 
\begin{equation}
\mathbfit{U}_n = n\mathbfit{U}_1 = n\ \!\frac{d}{\lambda} =  n\ \!\frac{d}{\lambda_0}
\ \!\frac{\lambda_0}{\lambda}
\end{equation}
The visibility, obtained as a Fourier sum in the simulation and given
in equation~\ref{eqn:modelvis}, simplifies to
\begin{equation}
\mathbfit{M}(\mathbfit{U}, \nu) = \sum\limits_{\boldsymbol \theta}\ {I}({\boldsymbol \theta},
\nu)\ {A}({\boldsymbol \theta},
\nu)\ e^{-i 2\pi m \frac{nd}{\lambda_0} \frac{\lambda_0}{\lambda}}
\end{equation}
Since $\mathcal{V}(\mathbfit{U},\nu)$ represents an estimate of
$\mathbfit{M}(\mathbfit{U})$, the two-visibility correlation at $(\nu_i, \nu_j)$ for
this baseline becomes
\begin{equation}
\mathbfss{S}_2(\mathbfit{U}_n, \nu_i, \nu_j) \sim  \left|{I}(\boldsymbol \theta, \nu_0)\right|^2
\left|{A}(\boldsymbol \theta, \nu_0)\right|^2 e^{-i 2\pi m \left(\frac{nd}{\lambda_0}\right)\left(\frac{\nu_i-\nu_j}{\nu_0}\right)}
\end{equation}
assuming that the approximations 
\begin{equation}
\left|{I}(\boldsymbol \theta, \nu_0)\right|^2 =
{I}(\boldsymbol \theta, \nu_i){I}^\ast(\boldsymbol \theta, \nu_j)
\label{eqn:I-approx}
\end{equation}
and 
\begin{equation}
\left|{A}(\boldsymbol \theta, \nu_0)\right|^2 =
{A}(\boldsymbol \theta, \nu_i){A}^\ast(\boldsymbol \theta, \nu_j)
\label{eqn:A-approx}
\end{equation}
are reasonable, but in general not strictly true.

The intrinsic frequency dependence of the sky signal can be disentangled from 
the chromatic instrument response by varying $\alpha$ and $\gamma$ over a range
of values. 
We observe the resulting spectral behaviour
$\mathbfss{S}_2$ for a single Gaussian realization of the diffuse
foreground. Figure~\ref{fig:V2_vs_beta} shows $\mathbfss{S}_2(\Delta\nu=0)$
normalised to its value at $\nu = 326.5$ MHz, for the longest baseline, where $\alpha$ is allowed to take three different
values: 0.0, 1.0 and 2.0 (the three curves in each panel) for each $\gamma=0, 1, 2$. 
The spatial index $\gamma$ exerts a spectral dependence through $\nu^{-\gamma}$,
and the spectral index $\alpha$ through $\nu^{2(\alpha-2)}$.
Therefore for $\gamma = 0$ and $\alpha=2$ (spectral index in temperature units), all
the spectral behaviour that manifests in the curve is indeed being caused by (a)
the baselines $\mathbfit{U}$
changing with frequency and (b) the primary beam shrinking
with increasing frequency. The strikingly similar curves for different values
of $\alpha$ and $\gamma$ point to their effect being sub-dominant.
Therefore, the bulk of the
spectral trend of $\mathbfss{S}_2$ in a single foreground realisation comes from the
chromatic response of the instrument(see also \citealt{Vedantham2012,Thyagarajan2013}).

\begin{figure*}
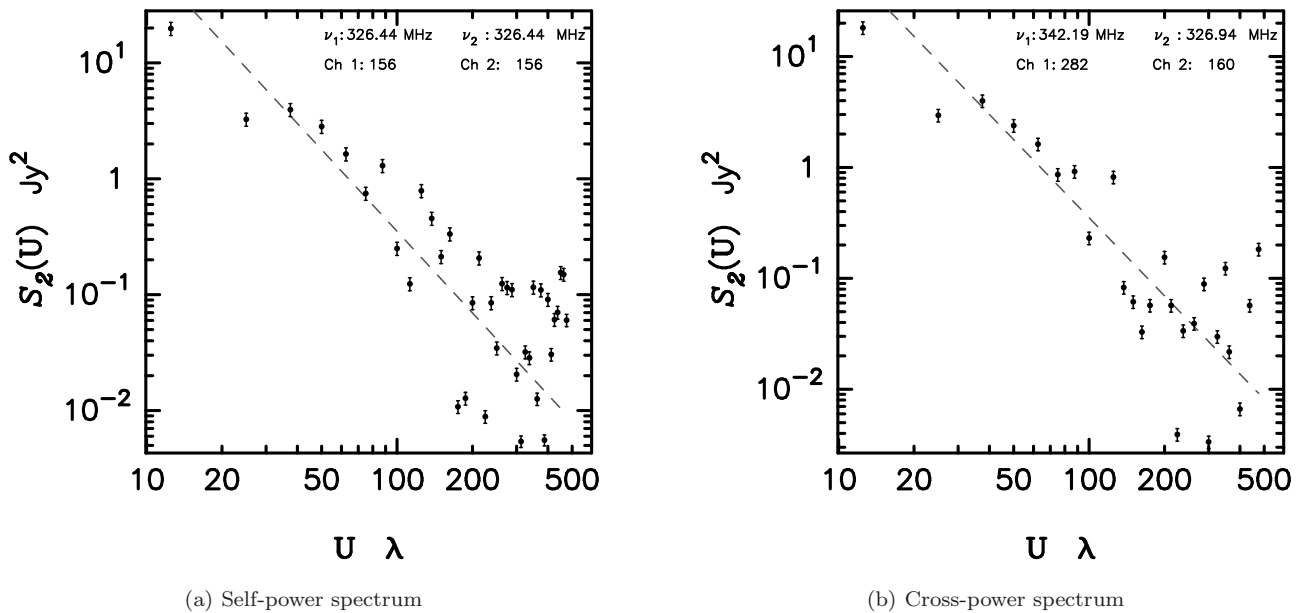

\begin{center}
\begin{minipage}{190mm}
\subfigure[Self-power spectrum]{\label{subfig:self-error}\includegraphics[scale=0.4,
  angle=270]{V2Corr.ch156-ch156.nu1_326.44_MHz-nu2_326.44_MHz.eps}}
\hskip 12mm
\subfigure[Cross-power spectrum]{\label{subfig:cross-error}\includegraphics[scale=0.4,
  angle=270]{V2Corr.ch282-ch160.nu1_342.19_MHz-nu2_326.94_MHz.eps}}
\end{minipage}
\end{center}
\caption{ The panels (a) and (b) respectively show the ``self-power
  spectrum'' and ``cross-power spectrum'' at different values of
  $(\nu_i, \nu_j)$ by sampling through the $\mathbfss{S}_2$
  cube. The error bars, drawn from the same co-ordinates of the RMS
  cube, correspond to 60 seconds of integration with a system
  temperature of $T_\mathrm{sys} = 150$ K. The dashed straight
  line is the input angular power spectrum of the
  Galactic diffuse synchrotron foreground. The scatter in the
  estimated power spectrum is due to the stochastic nature of a single realization of the
  foreground.}
\label{fig:V2_err_bar}
\end{figure*}
\subsection{Error on the estimator}
Having studied the visibility correlation estimator through noise-free
simulations, we could now turn our attention to the effects of noise.
The estimator $\mathbfss{S}_2$ has the dimensions of variance. For the error on
$\mathbfss{S}_2$ we are hence interested
in the variance of the variance. 
In the simulations described above, the visibility correlation matrix $\mathbfss{S}_2$
is computed at every one-second interval. For an $N$-second observation,
therefore, $N$ such matrices are available for each baseline, from which the mean matrix
and the RMS matrix for each baseline can be computed.

The estimator $\mathbfss{S}_2$ and the error matrix $\Delta\mathbfss{S}_2$ are
computed now for a single realisation of the Galactic diffuse foreground,
with the system temperature of 150 K \citep{Selvanayagam1993} and a 60-second observation.  This is
equivalent to deriving the error bars from as many independent
realisations (60) of the noise, given a single realisation of the foreground.
Figure~\ref{fig:V2_err_bar} shows the recovered power spectrum by sampling through the
mean and RMS cubes, $\mathbfss{S}_2$ and $\Delta\mathbfss{S}_2$, at the same $\left(\nu_i, \nu_j\right)$ co-ordinates:
the power spectrum is obtained from the mean $\mathbfss{S}_2$ matrix and the error bars
from the RMS $\Delta\mathbfss{S}_2$ matrix. Figure~\ref{subfig:self-error} shows the ``self'' power spectrum with the error bars
where $\nu_i = \nu_j =$326.44 MHz, and Figure~\ref{subfig:cross-error} shows the
``cross'' power spectrum with error bars, at $\nu_i= $342.19 MHz and $\nu_j = $326.94
MHz. The dashed straight line through the plots is not a fit, but it
is the input analytical power spectrum used for simulating the diffuse Galactic
foreground. The $\mathbfit{U}$ values on the $x$-axis are computed at the central frequency 326.5
MHz, as there is no other meaningful way to represent $\mathbfit{U}$ at a pair of frequencies
$(\nu_i, \nu_j)$ at which these spectra have been extracted.
Figure~\ref{fig:V2_err_bar} is an important result purely from the
perspective of foregrounds: a 60-second integration has resulted in a $10\sigma$
detection of the diffuse Galactic foregrounds, under the implicit assumption that the
point source foregrounds have been modelled out, or are sub-dominant as would
be the case for the short baselines of OWFA. This is a very optimistic scenario if
we are interested in the statistics of the diffuse
foregrounds.

\section{Summary}
In this paper we have briefly described the upcoming Ooty Wide Field Array
interferometer, one of whose goals is to detect the large scale structure in
redshifted 21-cm emission from $z \sim 3.35$. A detailed simulation of the
instrument and the foregrounds has been carried out. The simulations incorporate
a fully chromatic instrument model and hence lead to conservative estimates for
the expected systematics. We have recast the visibility correlation estimator,
previously applied to GMRT observations, for OWFA. We find that the foregrounds manifest with very smooth
spectra in the estimator, and that fitting them out with polynomials still
leaves systematic residuals that are 1-2 orders stronger than the cosmological
signal. Besides, the chromatic response of the instrument gives rise to
dominant spectral features in the bandpass of the baselines. The short baselines
make OWFA a sensitive instrument for rapidly characterising the diffuse
foregrounds. In a future paper, we will address further suppression of the
foreground residuals.

\section{Acknowledgements}
VRM thanks CTS, IIT Kharagpur, and SC thanks NCRA for
hosting them during a part of this work. SC is supported by a University Grants Commission Research
Fellowship. The Ooty Radio Telescope is located at Radio Astronomy Centre,
Udhagamandalam and operated by the National Centre for Radio
Astrophysics, Tata Institute of Fundamental Research.


\bibliographystyle{mnras}
\bibliography{mylist} % if your bibtex file is called example.bib

% Don't change these lines
\bsp	% typesetting comment
\label{lastpage}
\end{document}